\documentclass[onecolumn, draftclsnofoot, 12pt]{IEEEtran}
\usepackage{amsmath}
\usepackage{cite}
\usepackage{amssymb}
\usepackage{amsfonts}
\usepackage{algorithm}
\usepackage{dsfont}
\usepackage{graphicx}
\usepackage{epsfig}
\usepackage{subfigure}
\usepackage{psfrag}
\usepackage{xcolor}
\usepackage{url}
\usepackage[colorlinks,linkcolor=black,urlcolor=black,anchorcolor=black,citecolor=black,hyperfootnotes=true]{hyperref}

\newtheorem{lemma}{Lemma}
\newtheorem{theorem}{Theorem}

\newcommand{\mv}[1]{\mbox{\boldmath{$ #1 $}}}
\allowdisplaybreaks[4]
\title{Exploiting Temporal Side Information in Massive IoT Connectivity}
\author{Qipeng Wang, Liang Liu, Shuowen Zhang, Francis C.M. Lau
\thanks{The authors are with the Department of Electronic and Information Engineering, The Hong Kong Polytechnic University, Hong Kong, China (e-mails: qipeng.wang@connect.polyu.hk, \{liang-eie.liu, shuowen.zhang, francis-cm.lau\}@polyu.edu.hk).}
\thanks{This paper was presented in part at the IEEE International
Symposium on Information Theory (ISIT), July 2021 \cite{conference}.}}

\begin{document}
\maketitle \thispagestyle{empty} \vspace{-0.3in}

\begin{abstract}\label{sec_abstr}
This paper considers the joint device activity detection and channel estimation problem in a massive Internet of Things (IoT) connectivity system, where a large number of IoT devices exist but merely a random subset of them become active for short-packet transmission in each coherence block. In particular, we propose to leverage the \emph{temporal correlation} in device activity, e.g., a device active in the previous coherence block is more likely to be still active in the current coherence block, to improve the detection and estimation performance. However, it is challenging to utilize this temporal correlation as side information (SI), which relies on the knowledge about the exact statistical relation between the estimated activity pattern for the previous coherence block (which may be imperfect with unknown error) and the true activity pattern in the current coherence block. To tackle this challenge, we establish a novel \emph{SI-aided multiple measurement vector approximate message passing (MMV-AMP) framework}. Specifically, thanks to the \emph{state evolution} of the MMV-AMP algorithm, the correlation between the activity pattern estimated by the MMV-AMP algorithm in the previous coherence block and the real activity pattern in the current coherence block is quantified explicitly. Based on the well-defined temporal correlation, we further manage to embed this useful SI into the denoiser design under the MMV-AMP framework. Specifically, the SI-based soft-thresholding denoisers with binary thresholds and the SI-based minimum mean-squared error (MMSE) denoisers are characterized for the cases without and with the knowledge of the channel distribution, respectively. Numerical results are given to show the significant gain in device activity detection and channel estimation performance brought by our proposed SI-aided MMV-AMP framework.
\end{abstract}

\begin{IEEEkeywords}
Massive connectivity, approximate message passing (AMP), device activity detection, grant-free random access, temporal correlation, side information (SI).
\end{IEEEkeywords}

\section{Introduction}\label{sec_intro}

\subsection{Motivation}\label{sub_sec_mtv}
In a typical massive Internet of Things (IoT) connectivity system, one base station (BS) is expected to support $10^4$ to $10^6$ low-cost devices \cite{limit}. Due to the limited battery life, the low-cost IoT devices are generally designed to stay in silence for a long period to save energy and become active when triggered by external events, which leads to a sporadic traffic pattern \cite{mMTC1, survey}. In addition, the IoT traffic is usually with a stringent delay requirement such that the controllers can take actions in real time based on the sensing data. To reduce the access delay, the grant-free random access scheme \cite{mMTC2} has recently attracted a lot of attention, where the active devices can directly send the data to the BS without waiting for the permission from it. Therefore, for enabling the grant-free random access in massive IoT connectivity systems with sporadic and low-latency traffic, it is of paramount importance to investigate the strategies that can detect the device activity and/or estimate the corresponding channels in a fast and accurate manner.

In the literature, it has been shown that the joint device activity detection and channel estimation problem can be formulated as a sparse signal recovery problem and solved utilizing the compressed sensing techniques, thanks to the sparse device activity \cite{mMTC2,mMTC3}. In particular, under the framework of multiple measurement vector approximate message passing (MMV-AMP) \cite{AMP, MMV_AMP}, it has been shown in \cite{mMTC2} that the activity detection error probability decreases significantly as the number of antennas at the BS increases. Such an exciting result arises from exploiting the \emph{spatial correlation} in the device activities: if one device is active for one antenna, it is also active for all the other antennas, and the measurements at the large number of antennas can be jointly processed for improving the device activity detection accuracy. However, this theoretical performance gain is achieved at the cost of high computational complexity for processing a large number of measurements at the antennas, which is especially prohibitive for massive connectivity systems where the number of devices is already very large. Therefore, a natural question is: if only a small number of antennas are utilized to reduce the computational complexity, is it still possible to achieve high-quality device activity detection and channel estimation? This paper provides an affirm answer to the above question. Specifically, besides the spatial correlation, we identify that the \emph{temporal correlation} in the device activities among consecutive coherence blocks can also be exploited to improve the device activity detection and channel estimation performance. Such temporal correlation typically exists in practice due to the temporal correlation of the events triggering the device activity. For example, if a device (e.g., sensor) is activated at one moment due to some abnormal events, then this device is more likely to be still activated by this event in the near future. Thus, the estimation result in the previous coherence block can be leveraged as the side information (SI) for assisting the activity detection and channel estimation in the current coherence block. In this paper, we aim to establish a new SI-aided MMV-AMP framework to fully take advantage of the temporally-correlated device activity as the SI.

\subsection{Prior Work}\label{sub_sec_prwk}

In the literature, the device activity detection and/or channel estimation problem under the grant-free random access scheme has been widely studied for massive IoT connectivity.

\subsubsection{Compressed sensing-based approach for grant-free massive IoT connectivity}Thanks to the sparse device activity, the joint device activity and channel estimation problem can be cast as a compressed sensing problem. Along this line, the AMP algorithm \cite{AMP} was utilized for device activity detection and channel estimation in some early works \cite{mMTC2, mMTC3, mMTC4}, where the state evolution played a vital role for theoretically analyzing the performance of the AMP algorithm. Particularly, \cite{mMTC2} theoretically proved that with the MMV-AMP algorithm, the activity detection error goes to zero when the number of antennas at the BS goes to infinity, indicating that the massive multiple-input multiple-output (MIMO) technique is a good fit for massive IoT connectivity. The AMP framework was also generalized to other related areas. For example, the generalized MMV-AMP algorithm was considered in \cite{mMTC5} in a broadband IoT setup; \cite{mMTC_amp1} designed a transmission control scheme such that the user activity is more sparse to improve the AMP performance; \cite{mMTC_amp2} used the AMP algorithm for the scenario where the data is transmitted without previous preamble signals such that the data detection is performed together with the device activity detection. Besides the AMP algorithm, other compressed sensing techniques are also employed for device activity detection and channel estimation in massive IoT systems, including the learning technique \cite{mMTC_sbl1, mMTC_sbl2, mMTC_dplearning}, the dimension reduction technique \cite{mMTC_dr}, and the Reed-Muller detection technique \cite{mMTC_rmsq}.

In addition to the above standard compressed sensing framework, various SI exists in practice that can provide useful information to improve the sparse signal recovery performance. For example, when the partial knowledge about the support of the sparse signal to be estimated is available, it was shown in \cite{weighted_LASSO1, weighted_LASSO2} that the weighted $l_1$ minimization technique leads to an improved performance, where the weights associated with the penalty of the elements that are more likely to be zero are set to be larger values. On the other hand, the temporal correlation in the signals to be estimated, e.g., an IoT device in the previous coherence block will be more likely to be activated by the same event in the current coherence block, can also be utilized in compressed sensing algorithms \cite{SI-AMP, DCS-AMP}. Specifically, \cite{DCS-AMP} developed a Turbo extension of the AMP algorithm based on the idea of factor graph, where the SI is embedded into the factor graph models. However, this approach is not Bayesian-optimal and new factor graphs need to be craft carefully for each new signal model with high complexity. To simplify the process for utilizing the SI, \cite{SI-AMP} further proposed to incorporate the SI into the minimum mean-squared error (MMSE) denoiser design under the single measurement vector (SMV) AMP framework when the signal distribution information is known. However, in a multi-antenna communication system, how to embed the temporal SI into the MMSE denoiser design under the MMV-AMP framework is still an open problem. Moreover, if the channel distribution information is unknown, it is crucial to study how to utilize the temporal SI in the AMP algorithm, which is missing in the literature.

\subsubsection{Other approaches for grant-free massive IoT connectivity}

Other than the above compressed sensing-based approach, there are other strategies for grant-free massive IoT connectivity, including the covariance-based approach and the unsourced random access based approach. First, in the case when the channel estimation is not necessary, e.g., each device merely transmits a few bits which can be embedded into its preamble selection pattern, it was shown in \cite{cov1, cov2, cov3} that the minimum preamble sequence length for device activity detection can be significantly shortened by the covariance-based approach. However, if each device needs to transmit more bits such that their channels need to be estimated similar to the conventional data transmission, the above approach cannot be applied. On the other hand, the unsourced random access based approach was proposed in \cite{uscd1} and widely studied in \cite{uscd2, uscd3}. Under this approach, each device employs the same codebook, and the task of the decoder is to recover the list of transmitted messages irrespective of the identity of the devices because the device identifier can be embedded in the data. However, the decoding complexity under the unsourced random access based approach is extremely high.

\subsection{Main Contributions}\label{sub_sec_cntr}

This paper investigates the MMV-AMP algorithm for the joint device activity detection and channel estimation problem in a massive IoT connectivity system, where the device activities are temporally correlated. Specifically, two application scenarios are considered. In the first scenario, the distributions of the channels between the devices and the BS are assumed to be \emph{unknown}, e.g., when the devices are underground or underwater sensors for which an accurate channel model is difficult to obtain. In contrast, in the second scenario, the channel distribution information is assumed to be perfectly \emph{known}, e.g., for sensors in smart home and smart factories. For both scenarios, this paper aims to fully exploit the temporal correlation in the device activity for achieving better detection and estimation performance with a small number of antennas (thus with low computational complexity). The main contributions of this paper are summarized as follows.

\begin{itemize}
    \item First, we propose a novel \emph{SI-aided MMV-AMP framework} to incorporate the temporal SI into the MMV-AMP framework. Specifically, in each coherence block, we aim to jointly utilize the estimation result in the previous coherence block and the temporal activity correlation to improve the activity detection and channel estimation performance. However, this is a challenging task since the estimation result in the previous coherence block is generally imperfect, thus cannot be directly utilized as the SI. Thanks to the state evolution of the MMV-AMP algorithm, we model the useful SI theoretically and characterize the exact statistical relationship between the SI in the previous coherence block and the real effective channels in the current coherence block, which will serve as the basis of the denoiser designs for both cases with unknown or known channel distribution information.

    \item Next, for the case without knowledge of the channel distribution information, we design the \emph{SI-based soft-thresholding denoisers} under the MMV-AMP framework. By formulating the SI-based sparse signal recovery problem as a weighted group Least Absolute Shrinkage and Selection Operator (LASSO) problem, we first derive the closed-form SI-based soft-thresholding denoisers with a binary threshold - the threshold in the denoisers in the current coherence block depends on whether a device is detected to be active or inactive in the previous coherence block. This is in sharp contrast to the conventional soft-thresholding denoisers without SI where the threshold is not binary. Inspired by the minimax approach that is widely used for the threshold design in soft-thresholding denoisers without channel distribution information, we further design the binary threshold to optimize the (approximate) worst-case estimation mean-squared error (MSE) with the (approximate) least-favorable channel distribution. To our best knowledge, this is the first result to design the SI-based binary threshold for soft-thresholding denoisers in the literature.

    \item Furthermore, for the case with knowledge of the channel distribution information, we design the \emph{SI-based MMSE denoisers} under the MMV-AMP framework. Based on the exact statistical relationship between the SI in the previous coherence block and the device effective channel in the current coherence block, we manage to characterize the closed-form SI-based MMSE denoisers by minimizing the conditional MSE for estimating the device effective channels. It is worth noting that our SI-based MMSE denoisers are Bayesian-optimal and designed without the need of crafting complex factor graphs for every new signals as compared to \cite{DCS-AMP}. Moreover, different from \cite{DCS-AMP} and \cite{SI-AMP}, our denoisers can also work for the more general MMV-AMP framework in multi-antenna systems.
\end{itemize}

\subsection{Organization}\label{sub_sec_org}
The rest of the paper is organized as follows. Section \ref{sec_sys} introduces the system model. Section \ref{sec_si_amp} establishes the SI-aided MMV-AMP framework. Section \ref{sec_unknown} and Section \ref{sec_known} present the SI-based denoiser design with unknown channel distribution information or with known channel distribution information, respectively. Numerical results are provided in Section \ref{sec_num}. Finally, conclusions are drawn in Section \ref{sec_conclusion}.

\emph{Notations}: Throughout the paper, scalars are denoted by lower-case letters, vectors by bold-face lower-case letters, and matrices by bold-face upper-case letters. The identity matrix and the all-zero matrix of appropriate dimensions are denoted as $\boldsymbol{I}$ and $\mv{0}$. For a full rank matrix $\boldsymbol{M}$, $\boldsymbol{M}^{-1}$ denotes its inverse. For a matrix $\boldsymbol{M}$, $\boldsymbol{M}^H$ and $\boldsymbol{M}^T$ denote its conjugate transpose and transpose, respectively. The expectation operator is denoted as $\mathop{\mathbb{E}}[\cdot]$. The probability of an event is denoted as $Pr(\cdot)$. The distribution of a circularly symmetric complex Guassian random vector with mean $\mv{x}$ and covariance matrix $\mv{\Sigma}$ is denoted as $\mathcal{CN}(\boldsymbol{x},\boldsymbol{\Sigma})$. $\sim$ denotes distributed as. $\|\cdot\|_0$ and $\|\cdot\|$ denote the $l_0$ norm and $l_2$ norm, respectively.

\section{System Model}\label{sec_sys}

\subsection{Baseband Model}\label{sub_sec_bas}
This paper considers the uplink communication in a massive IoT connectivity system consisting of one BS equipped with $M$ antennas and $N$ single-antenna IoT devices. We assume quasi-static block-fading channels, in which all user channels remain approximately constant in each coherence block, but vary independently from block to block. Let $J$ denote the number of consecutive coherence blocks considered in this work. In each coherence block $j$, the channel from device $n$ to the $m$-th antenna of the BS is denoted by ${h}_{n,m}^{(j)}$, whose distribution is denoted by $\mu_{n,m}^{(j)}$, i.e., ${h}_{n,m}^{(j)}\sim\mu_{n,m}^{(j)}$, $\forall n, m, j$. It is assumed that for each $n$, ${h}_{n,m}^{(j)}$'s are independent and identically distributed (i.i.d.) over $m$ and $j$, i.e., ${\mu}_{n,m}^{(j)}=\mu_{n},\forall m,j$. For convenience, define $\boldsymbol{h}_n^{(j)}=[h_{n,1}^{(j)},\cdots,h_{n,M}^{(j)}]^{T}\in\mathbb{C}^{M\times1}, \forall n,j$. Then, each channel $\boldsymbol{h}_n^{(j)}$ is distributed according to the product measure $\boldsymbol{\mu}_{n,M}=\mu_n\times\cdots\times\mu_n\in \mathcal{G}_{M}$, where $\mathcal{G}_{M}$ is the family of probability measures over $\mathbb{C}^{M\times 1}$. In practice, $\boldsymbol{\mu}_{n,M}$'s may be unknown or known to the BS. We will study these two cases in Sections \ref{sec_unknown} and \ref{sec_known}, respectively.

Due to the sporadic data traffic in IoT networks, only a small set of devices become active in each coherence block. We define the device activity indicator functions as follows:

\begin{equation}\label{indicator}
     \delta_n^{\left(j\right)}=
        \begin{cases}
        1,& \text{if device $n$ is active in coherence block} ~ j,\\
        0,& \text{otherwise},
        \end{cases} \quad \forall n,j,
    \end{equation}so that $\delta_n^{\left(j\right)}$ is a Bernoulli random variable with
\begin{equation}\label{p_active}
    Pr(\delta_n^{\left(j\right)}=1)=\lambda, ~ Pr(\delta_n^{\left(j\right)}=0)=1-\lambda, \quad \forall n, j.
\end{equation}

In this paper, we consider the grant-free random access scheme \cite{mMTC1} in our interested IoT system, where at the beginning of each coherence block, the active devices transmit their pilot sequences to the BS to perform joint device activity detection and channel estimation. Let $\boldsymbol{s}_n=[{s}_{n,1}, \ldots, {s}_{n,L}]^T\in \mathbb{C}^{L\times 1}$ denote the pilot sequence with length $L$ assigned to device $n$, $\forall n$. Similar to \cite{mMTC1,mMTC2,mMTC3}, it is assumed that all the entries in $\boldsymbol{s}_n$ are generated according to the i.i.d. complex Gaussian distribution with zero mean and variance $1/L$, $\forall n$. Then, the received signal at the BS in coherence block $j$ is expressed as

\begin{equation}\label{basebandmodel}
    \boldsymbol{Y}^{\left(j\right)}= \sum_{n=1}^{N} \delta_n^{\left(j\right)}\boldsymbol{s}_n(\boldsymbol{h}_n^{\left(j\right)})^T+\boldsymbol{Z}^{\left(j\right)}=\boldsymbol{S}\boldsymbol{X}^{\left(j\right)}+\boldsymbol{Z}^{\left(j\right)}, \quad \forall j,
\end{equation}
where $\boldsymbol{Z}^{\left(j\right)} \in \mathbb{C}^{L\times M} \sim \mathcal{CN}(\boldsymbol{0},\sigma_z^2\boldsymbol{I})$ is the noise at the BS in coherence block $j$ whose variance $\sigma_z^2$ depends on the background noise power normalized by user transmit power, $\boldsymbol{S}=[\boldsymbol{s}_1,\ldots,\boldsymbol{s}_N]\in\mathbb{C}^{L\times N}$, and $\boldsymbol{X}^{\left(j\right)}=[\boldsymbol{x}_1^{\left(j\right)}, \ldots,\boldsymbol{x}_N^{\left(j\right)}]^T \in\mathbb{C}^{N\times M}$ with $\boldsymbol{x}_n^{\left(j\right)}=\delta_n^{\left(j\right)}\boldsymbol{h}_n^{\left(j\right)}$ denoting the effective channel of device $n$ in coherence block $j$, $\forall n,j$. According to (\ref{p_active}), the effective channel $\boldsymbol{x}_n^{\left(j\right)}$ follows the distribution $\boldsymbol{\nu}_{n,M}\in \mathcal{F}_{M,\lambda}$, where $\mathcal{F}_{M,\lambda}\equiv\{\boldsymbol{\nu}_{n}\mid  \mathbb{E}_{\boldsymbol{x}_n^{(j)}\sim \boldsymbol{\nu}_{n,M}}[\|\boldsymbol{x}_n^{\left(j\right)}\|_{0}]\leq  M\lambda\}$ with $\|\boldsymbol{x}_n^{\left(j\right)}\|_{0}$ indicating the number of non-zero elements in $\boldsymbol{x}_n^{(j)}$, $\forall n, j$. Note that if device $n$ is active, $\boldsymbol{\nu}_{n,M}$ will reduce to $\boldsymbol{\mu}_{n,M}$, i.e., the distribution of $\boldsymbol{h}_n$. Depending on the knowledge of channel distribution, the distribution of the effective channels may be unknown or known to the BS. In each coherence block $j$, the job of the BS is to jointly detect the active devices and estimate their channels by estimating $\boldsymbol{X}^{\left(j\right)}$ based on its received signal $\boldsymbol{Y}^{\left(j\right)}$ and its knowledge of the device pilots $\boldsymbol{S}$, without or with information about the distribution $\boldsymbol{\nu}_{n,M}$'s.

\subsection{Temporally-Correlated Device Activity Model}\label{sub_sec_tem_cor}
This paper considers the case of temporally-correlated device activity, which is modeled by a Markov chain with the following transition probabilities:

\begin{equation}\label{p_transition}
\begin{aligned}
    Pr(\delta_n^{\left(j\right)} = 1\mid \delta_n^{\left(j-1\right)}=1)&=\alpha,\\
    Pr(\delta_n^{\left(j\right)} = 0\mid \delta_n^{\left(j-1\right)}=1)&=1-\alpha,\\
    Pr(\delta_n^{\left(j\right)}= 1\mid \delta_n^{\left(j-1\right)}=0)&=\beta,\\
    Pr(\delta_n^{\left(j\right)} = 0\mid \delta_n^{\left(j-1\right)}=0)&=1-\beta,
\end{aligned}\quad \forall n,j.
\end{equation}In other words, if device $n$ is active in coherence block $j-1$, then with probability $\alpha\in[\lambda,1)$, it is still active in coherence block $j$; if device $n$ is inactive in coherence block $j-1$, then with probability $\beta\in(0,\lambda]$, it is active in coherence block $j$.\footnote{In practice, the values of $\alpha$ and $\beta$ can be learned by the method proposed in \cite{par_learn}.} Given the above temporal correlation, we have the following four cases to model each device's activity over two consecutive coherence blocks $j-1$ and $j$:

\emph{Case 1}: A device is active for both coherence blocks $j-1$ and $j$, i.e., $\boldsymbol{x}_n^{\left(j-1\right)}=\boldsymbol{h}_n^{\left(j-1\right)}$ and $\boldsymbol{x}_n^{\left(j\right)}=\boldsymbol{h}_n^{\left(j\right)}$, with probability $\alpha\lambda$.

\emph{Case 2}: A device is active in coherence block $j-1$, but becomes inactive in coherence block $j$, i.e., $\boldsymbol{x}_n^{\left(j-1\right)}=\boldsymbol{h}_n^{\left(j-1\right)}$ and $\boldsymbol{x}_n^{\left(j\right)}=\boldsymbol{0}$, with probability $(1-\alpha)\lambda$.

\emph{Case 3}: A device is inactive in coherence block $j-1$, but becomes active in coherence block $j$, i.e., $\boldsymbol{x}_n^{\left(j-1\right)}=\boldsymbol{0}$ and $\boldsymbol{x}_n^{\left(j\right)}=\boldsymbol{h}_n^{\left(j\right)}$, with probability $\beta(1-\lambda)$.

\emph{Case 4}: A device is inactive for both coherence blocks $j-1$ and $j$, i.e., $\boldsymbol{x}_n^{\left(j-1\right)}=\boldsymbol{0}$ and $\boldsymbol{x}_n^{\left(j\right)}=\boldsymbol{0}$, with probability $(1-\beta)(1-\lambda)$.

Similar to \cite{DCS-AMP} and \cite{SI-AMP}, we assume that each Markov chain operates in steady-state such that the probability that a device becomes active is $\lambda$ over all the $J$ coherence blocks, i.e., (\ref{p_active}). Under this condition, the relation between $\alpha$ and $\beta$ is given by
\begin{equation}
    \alpha\lambda+\beta(1-\lambda)=\lambda.
\end{equation}Due to this relation, the Markov chains are completely characterized by two parameters $\lambda$ and $\alpha$.

Under the temporal correlation modeled by (\ref{p_transition}), we should not detect the device activity over consecutive coherence blocks in an independent manner as in \cite{mMTC2, mMTC3}, since the device activity in the previous coherence block can provide SI for improving the detection and estimation accuracy in the current coherence block. However, in each coherence block $j$, only an imperfect estimation of $\boldsymbol{x}_n^{(j-1)}$, $\forall n$, for the previous coherence block $j-1$, denoted by $\hat{\boldsymbol{x}}_n^{(j-1)}$, $\forall n$, is available at the BS. Despite the temporal correlation shown in (\ref{p_transition}), it is non-trivial to model a precise statistical relation between $\boldsymbol{x}_n^{(j)}$ and $\hat{\boldsymbol{x}}_n^{(j-1)}$, $\forall n$, since the connection between $\boldsymbol{x}_n^{(j-1)}$ and $\hat{\boldsymbol{x}}_n^{(j-1)}$, $\forall n$, is in general unknown. Without knowing such a statistical relation, it is possible that the imperfect estimation in the previous coherence block provides wrong SI for the estimation in the current coherence block, which may even degrade the system performance. This motivates us to pursue a systematic approach that is able to leverage the SI to improve the average performance of the activity detection and channel estimation.

Note that in the case without using SI, \cite{mMTC2} and \cite{mMTC3} showed that the estimation of $\boldsymbol{X}^{\left(j\right)}$ based on (\ref{basebandmodel}) is a compressed sensing problem, since many rows in $\boldsymbol{X}^{\left(j\right)}$ are zero vectors due to the sparse device activity. Moreover, the MMV-AMP algorithm has been used to estimate the row-sparse matrix $\boldsymbol{X}^{\left(j\right)}$ in each coherence block. In the rest of this paper, under the framework of MMV-AMP, we study the statistical relation between $\hat{\boldsymbol{x}}_n^{(j-1)}$'s and $\boldsymbol{x}_n^{(j)}$'s in adjacent blocks and show how this relation can be utilized to establish an efficient SI-aided MMV-AMP framework.

\section{SI-aided MMV-AMP Framework}\label{sec_si_amp}

In this section, we establish the SI-aided MMV-AMP framework. Under this framework, we will introduce what SI should be used and how to utilize it to improve the performance.

\subsection{SI-Aided MMV-AMP Framework}\label{sub_sec_alg}

In coherence block $j$, the SI-aided MMV-AMP algorithm will generate an estimation of $\boldsymbol{X}^{(j)}$, denoted by $\hat{\boldsymbol{X}}^{(j)}=[\hat{\boldsymbol{x}}_1^{(j)},\ldots,\hat{\boldsymbol{x}}_N^{(j)}]^T$, based on the signal received in the current coherence block as shown in (\ref{basebandmodel}) and the estimation made by SI-aided MMV-AMP algorithm in the previous coherence block, i.e., $\hat{\boldsymbol{X}}^{(j-1)}$. Specifically, in coherence block $j$, the SI-aided MMV-AMP algorithm starts from $\boldsymbol{X}_0^{\left(j\right)}=\boldsymbol{0}$ , $\boldsymbol{R}_0^{\left(j\right)}=\boldsymbol{Y}^{\left(j\right)}$, and then iterates as follows:
\begin{align}
    &\boldsymbol{x}_{n, t+1}^{\left(j\right)}= \eta_{n, t}^{\left(j\right)}\left ( \boldsymbol{x}_{n, t}^{\left(j\right)}+(\boldsymbol{R}_t^{\left(j\right)})^H \boldsymbol{s}_{n},  f_{n,j}\big(\boldsymbol{\hat{x}}_n^{(j-1)}\big)\right), \label{amp_x} \\
    &\boldsymbol{R}_{t+1}^{\left(j\right)}=  \boldsymbol{Y}^{\left(j\right)}- \boldsymbol{S}\boldsymbol{X}_{t+1}^{\left(j\right)} +\frac{N}{L} \boldsymbol{R}_{t}^{(j)} \left \langle  {\eta_{n,t}^{\left(j\right)}}' \left (\boldsymbol{x}_{n,t}^{\left(j\right)} + \big(\boldsymbol{R}_t^{\left(j\right)}\big)^H \boldsymbol{s}_{n}, f_{n,j} \big(\boldsymbol{\hat{x}}_n^{\left(j-1\right)}\big)\right) \right\rangle. \label{amp_r}
\end{align}
In (\ref{amp_x}) and (\ref{amp_r}), $t$ denotes the index of algorithm iteration starting from $0$, $\boldsymbol{X}_{t}^{\left(j\right)} = [\boldsymbol{x}_{1, t}^{\left(j\right)}, \ldots, \boldsymbol{x}_{N, t}^{\left(j\right)}]^T$ denotes the estimation of $\boldsymbol{X}^{\left(j\right)}$ at the $t$-th iteration of the SI-aided MMV-AMP algorithm, $f_{n,j}(\boldsymbol{\hat{x}}_n^{\left(j-1\right)})$ is a function of $\boldsymbol{\hat{x}}_n^{\left(j-1\right)}$ which is used as the SI for device $n$, $\boldsymbol{R}_t^{\left(j\right)}$ is the corresponding residual at iteration $t$, ${\eta}_{n, t}^{\left(j\right)} (\cdot, \diamond)\in\mathbb{C}^{M\times 1}$ is the denoising function for device $n$, ${\eta_{n, t}^{\left(j\right)}}' (\cdot, \diamond)$ is the first-order derivative of ${\eta}_{n, t}^{\left(j\right)}(\cdot, \diamond)$ with respect to the first variable $\cdot$, and $\left \langle \cdot  \right \rangle $ is the averaging operation over all entries of ${{\eta}_{n, t}^{\left(j\right)}}'( \cdot, \diamond )$. Let $\boldsymbol{X}_{\infty}^{\left(j\right)} = [\boldsymbol{x}_{1, \infty}^{\left(j\right)}, \ldots, \boldsymbol{x}_{N, \infty}^{\left(j\right)}]^T$ and $\boldsymbol{R}_{\infty}^{\left(j\right)}$ denote the estimation of $\boldsymbol{x}_{n}^{\left(j\right)}$ and the corresponding residual after the convergence of the SI-aided MMV-AMP algorithm in coherence block $j$. Then, we have $\hat{\boldsymbol{x}}_n^{\left(j\right)}=\boldsymbol{x}_{n,\infty}^{\left(j\right)}$, $\forall n,j$. Note that after the convergence of the SI-aided MMV-AMP algorithm in the $j$-th coherence block, the device activity detection is done by performing the log-likelihood ratio (LLR)-based detection to $\hat{\boldsymbol{x}}_n^{\left(j\right)}$, $\forall n$. Specifically, define $H_0$ and $H_1$ as the hypotheses that a device is inactive and active, respectively. Then, the LLR-based detection rule is \cite{mMTC3}:
\begin{equation}\label{llr_detection}
    \|\hat{\boldsymbol{x}}_n^{\left(j\right)}+\big(\boldsymbol{R}_\infty^{\left(j\right)}\big)^H \boldsymbol{s}_{n}\|\overset{H_1}{\underset{H_0}{\gtrless}}l,\quad \forall n,j,
\end{equation}
where $l$ is a common threshold for all the devices.

To summarize, under our proposed framework, we first implement the SI-aided MMV-AMP algorithm shown in (\ref{amp_x}) and (\ref{amp_r}) to estimate the user effective channels, and then apply the LLR-based approach shown in (\ref{llr_detection}) to detect the active devices. Since the LLR-based detectors are standard, in the rest of this paper, we focus on the open problem of how to embed the SI into the MMV-AMP algorithm design to improve the performance of the conventional MMV-AMP algorithms used in \cite{mMTC2}, \cite{mMTC3}. It is observed in (\ref{amp_x}) and (\ref{amp_r}) that there are two challenges in designing the SI-aided MMV-AMP algorithm: what SI, i.e., $f_{n,j}(\hat{\boldsymbol{x}}_n^{\left(j-1\right)})$, should be used and how to design the denoisers by utilizing the SI. In the following two subsections, we tackle the above two issues, respectively.

\subsection{Identifying SI From State Evolution}
Under the SI-aided MMV-AMP framework, there exists the state evolution in the asymptotic regime where $N,K,L\to \infty$ with fixed $N/L$ and $N/K$. Specifically, at each iteration $t$ of the AMP algorithm to estimate $\boldsymbol{X}^{(j)}$, $ \boldsymbol{x}_{n,t}^{\left(j\right)}+(\boldsymbol{R}_t^{\left(j\right)})^H \boldsymbol{s}_{n}$ is statistically equivalent to:
\begin{equation}\label{model}
    \boldsymbol{\tilde{x}}_{n,t}^{\left(j\right)} = \boldsymbol{x}_n^{\left(j\right)}+\big( \boldsymbol{\Sigma}_t^{\left(j\right)}\big)^\frac{1}{2}\boldsymbol{v}_n^{(j)},\quad \forall n,j,t,
\end{equation}where $\boldsymbol{v}_n^{(j)}\in\mathbb{C}^{M\times 1}\sim\mathcal{CN}(\boldsymbol{0},\boldsymbol{I})$ is the noise independent of $\boldsymbol{x}_n^{\left(j\right)}$ and $\boldsymbol{\Sigma}_t^{\left(j\right)}\in \mathbb{C}^{M\times M}$ is the \emph{state}.  Define a set of random vectors $\boldsymbol{X}_n^{\left(j\right)}\in\mathbb{C}^{M\times 1}$, $\boldsymbol{V}_n^{(j)}\in\mathbb{C}^{M\times 1}$, and $\boldsymbol{\hat{X}}_n^{\left(j-1\right)}\in\mathbb{C}^{M\times 1}$ which capture the distribution of $\boldsymbol{x}_n^{\left(j\right)}$, $\boldsymbol{v}_n^{(j)}$, and $\boldsymbol{\hat{x}}_n^{\left(j-1\right)}$, respectively, $\forall n,j$. Then, in the asymptotic regime where $N,K,L\to \infty$ with fixed $N/L$ and $N/K$, the state evolution is given by:
\begin{align}\label{state_evolution}
    \boldsymbol{\Sigma}_{t+1}^{\left(j\right)} &= \sigma_z^2\boldsymbol{I}+\frac{N}{L}\times\frac{1}{N}\sum_{n=1}^{N}\mathop{\mathbb{E}} \big[\big(\eta_{n,t}^{\left(j\right)}\big( \boldsymbol{X}_n^{\left(j\right)}+(  \boldsymbol{\Sigma}_t^{\left(j\right)} )^\frac{1}{2}\boldsymbol{V}_n^{(j)},f_{n,j} ( \boldsymbol{\hat{X}}_n^{\left(j-1\right)}) \big) \!-\!\boldsymbol{X}_n^{\left(j\right)} \big)\notag\\
    &\qquad\qquad\qquad\qquad\quad\quad\times \big( \eta_{n,t}^{\left(j\right)}\big(  \boldsymbol{X}_n^{\left(j\right)}+ (\boldsymbol{\Sigma}_t^{\left(j\right)})^\frac{1}{2}\boldsymbol{V}_n^{(j)}, f_{n,j}(\boldsymbol{\hat{X}}_n^{\left(j-1\right)}) \big) - \boldsymbol{X}_n^{\left(j\right)}\big)^{H} \big],\quad \forall j, t.
\end{align}
 Note that in coherence block $j$, we already have the estimation of $\boldsymbol{X}^{(j-1)}$, i.e., $\boldsymbol{\hat{x}}_n^{\left(j-1\right)}=\boldsymbol{x}_{n,\infty}^{\left(j-1\right)}$, $\forall n$. According to (\ref{model}), $\boldsymbol{\hat{x}}_{n}^{\left(j-1\right)}+(\boldsymbol{R}_\infty^{\left(j-1\right)})^H \boldsymbol{s}_{n}$ in the previous coherence block is statistically equivalent to
\begin{equation}\label{state_estimation}
    \boldsymbol{\tilde{x}}_{n,\infty}^{\left(j-1\right)}= \boldsymbol{x}_n^{\left(j-1\right)}+\big(\boldsymbol{\Sigma}_\infty^{\left(j-1\right)}\big)^{\frac{1}{2}}\boldsymbol{v}_n^{(j-1)}, ~~~ \forall n,j,t,
\end{equation}
where $\boldsymbol{\Sigma}_\infty^{\left(j-1\right)}$ denotes the state of the MMV-AMP algorithm shown in (\ref{state_evolution}) after it converges in coherence block $j-1$. It is worth noting that the correlation between the previous block's estimated effective channels plus residue, i.e., $\boldsymbol{\hat{x}}_{n}^{\left(j-1\right)}+(\boldsymbol{R}_\infty^{\left(j-1\right)})^H \boldsymbol{s}_{n}$'s, and true effective channels, i.e., $\boldsymbol{x}_n^{\left(j-1\right)}$'s, can be built based on (\ref{state_estimation}) as well as the statistical equivalence between $\boldsymbol{\hat{x}}_{n}^{\left(j-1\right)}+(\boldsymbol{R}_\infty^{\left(j-1\right)})^H \boldsymbol{s}_{n}$'s and $\boldsymbol{x}_n^{\left(j-1\right)}+(\boldsymbol{\Sigma}_\infty^{\left(j-1\right)})^{\frac{1}{2}}\boldsymbol{v}_n^{(j-1)}$'s, while the correlation between the previous coherence block's effective channels and the current coherence block's effective channels, i.e., $\boldsymbol{x}_n^{\left(j\right)}$'s, can be built based on (\ref{p_transition}). Thus, the correlation between $\boldsymbol{\hat{x}}_{n}^{\left(j-1\right)}+(\boldsymbol{R}_\infty^{\left(j-1\right)})^H\boldsymbol{s}_{n}$ and $\boldsymbol{x}_n^{\left(j\right)}$ can be built for all the devices and all the adjacent blocks. Moreover, after the convergence of the MMV-AMP in coherence block $j-1$, $\boldsymbol{\hat{{x}}}_n^{(j-1)}$ and $(\boldsymbol{R}_\infty^{\left(j-1\right)})^H \boldsymbol{s}_{n}$ can be obtained via (\ref{amp_x}) and (\ref{amp_r}), respectively. This motivates us to adopt the following SI to design the denoisers in the MMV-AMP algorithm:
\begin{equation}\label{SI}
    f_{n,j}\big(\boldsymbol{\hat{x}}_n^{\left(j-1\right)}\big)=\boldsymbol{\hat{x}}_n^{\left(j-1\right)}+\big(\!\boldsymbol{R}_\infty^{\left(j-1\right)}\big)^H \boldsymbol{s}_{n}, \quad\forall n,j.
\end{equation}
After identifying the SI, the next question is how to utilize the correlation between $\boldsymbol{\hat{x}}_n^{\left(j-1\right)}+(\boldsymbol{R}_\infty^{\left(j-1\right)})^H \boldsymbol{s}_{n}$'s and $\boldsymbol{x}_n^{\left(j\right)}$'s to design the denoisers in (\ref{amp_x}) for both the cases with unknown channel distribution information and known channel distribution information.

\subsection{Denoiser Design for SI-Aided MMV-AMP Framework}

Under the AMP framework, as long as the denoiser is a Lipschitz continuous function, the algorithm will converge. However, different denoisers will lead to very diverse performance. To achieve the best performance in the MMV-AMP algorithm, the denoisers should be carefully designed based on (\ref{model}) and the SI given in (\ref{SI}). If the distribution of $\boldsymbol{x}_n^{\left(j\right)}$'s, i.e., $\boldsymbol{\nu}_{n,M}$'s, is unknown in (\ref{model}), we can apply the group LASSO technique \cite{group_LASSO} to design the denoisers, similar to \cite{AMP_PT}. Specifically, the group LASSO problem for the linear model (\ref{model}) is characterized as
\begin{equation}\label{wglasso}
    \mathop{\text{minimize}}_{\eta_{n,t}^{\left(j\right)}\left (\cdot, \diamond \right )}\frac{1}{2}\!\left \|{\boldsymbol{\tilde{x}}}_{n,t}^{\left(j\right)}\!-\!\eta_{n, t}^{\left(j\right)}\!\big(\!\boldsymbol{\tilde{x}}_{n,t}^{\left(j\right)}, f_{n,j}\!\big(\boldsymbol{\hat{x}}_n^{\left(j\!-\!1\right)\!}\big)\!\big) \!\right \|^2\!+\theta_{n,t}^{\left(j\right)}\!\big(f_{n,j}\big(\!\boldsymbol{\hat{x}}_n^{\left(j-1\right)}\big)\!\big)\!\left \|\eta_{n, t}^{\left(j\right)}\!\big(\boldsymbol{\tilde{x}}_{n,t}^{\left(j\right)}, f_{n,j}\!\big(\boldsymbol{\hat{x}}_n^{\left(j-1\right)}\big)\!\big)\!\right \|, \quad\forall n, j, t,
\end{equation}
where $\theta_{n,t}^{\left(j\right)}(f_{n,j}(\boldsymbol{\hat{x}}_n^{\left(j-1\right)}))\!>\!0$ is a known parameter (which, however, needs to be carefully designed based on the SI) to control the sparsity of $\eta_{n, t}^{\left(j\right)}\!\big(\boldsymbol{\tilde{x}}_{n,t}^{\left(j\right)}, f_{n,j}\!\big(\boldsymbol{\hat{x}}_n^{\left(j-1\right)}\big)\!\big)$. Note that under our proposed SI-aided MMV-AMP framework, $\theta_{n,t}^{\left(j\right)}(f_{n,j}(\boldsymbol{\hat{x}}_n^{\left(j-1\right)}))$ is a function of the SI given in (\ref{SI}). Intuitively, if device $n$ in the previous coherence block is detected as active (or inactive), i.e., $ \| f_{n,j} (\boldsymbol{\hat{x}}_n^{\left(j-1\right)} )  \|> l$ (or $\| f_{n,j} (\boldsymbol{\hat{x}}_n^{\left(j-1\right)})  \|<l$), $\theta_{n,t}^{\left(j\right)}(f_{n,j}(\boldsymbol{\hat{x}}_n^{\left(j-1\right)}))$ should be set to be a smaller (or larger) value in the current coherence block such that $\boldsymbol{x}_n^{\left(j\right)}$ tends to be a non-zero (or zero) vector, i.e., device $n$ will be detected to be active (or inactive) with a higher probability. As a result, $\theta_{n,t}^{\left(j\right)}(f_{n,j}(\boldsymbol{\hat{x}}_n^{\left(j-1\right)}))$ can be defined as a binary threshold \cite{weighted_LASSO1} \cite{weighted_LASSO2}:
\begin{equation}\label{iii_threshold}
    \theta_{n,t}^{(j)}\big(f_{n,j}\big(\boldsymbol{\hat{x}}_n^{\left(j-1\right)}\big)\big)=\begin{cases}
        \theta_{1,n,t}^{(j)},  & \text{if}~  \| f_{n,j} \big(\boldsymbol{\hat{x}}_n^{\left(j-1\right)}\big ) \|>  l, \\
        \theta_{2,n,t}^{(j)},  & \text{if}~   \| f_{n,j} \big(\boldsymbol{\hat{x}}_n^{\left(j-1\right)}\big )  \|<  l,
    \end{cases} \quad\forall n, j, t,
\end{equation}
where $\theta_{1,n,t}^{(j)}\le\theta_{2,n,t}^{(j)}$. Given the above binary threshold $\theta_{n,t}^{\left(j\right)}(f_{n,j}(\boldsymbol{\hat{x}}_n^{\left(j-1\right)}))$, the closed-form solution to problem (\ref{wglasso}) is
\begin{equation}\label{sysm_si_st_denoiser}
    \eta_{n, t}^{\left(j\right)}\big(\boldsymbol{\tilde{x}}_{n,t}^{\left(j\right)}, \!f_{n,j}\!\big(\boldsymbol{\hat{x}}_n^{\left(j-1\right)}\big)\!\big)=\begin{cases}
    \!{\boldsymbol{\tilde{x}}}_{n,t}^{\left(j\right)}\!-\!\frac{\theta_{n,t}^{\left(j\right)}\!\big(\!f_{n,j}\!\big(\boldsymbol{\hat{x}}_n^{\left(j-1\right)}\big)\!\big) {\boldsymbol{\tilde{x}}}_{n,t}^{\left(j\right)}}{\|\tilde{\boldsymbol{x}}_{n,t}^{\left(j\right)}\|},  & \text{if}~  \!\|{\boldsymbol{\tilde{x}}}_{n,t}^{\left(j\right)}\|\!\ge\!\theta_{n,t}^{\left(j\right)}\big(f_{n,j}\!\big(\boldsymbol{\hat{x}}_n^{\left(j-1\right)}\big)\!\big)\!, \\
    \boldsymbol{0},  & \text{if}~  \!\|{\boldsymbol{\tilde{x}}}_{n,t}^{\left(j\right)}\|\!<\!\theta_{n,t}^{\left(j\right)}\big(f_{n,j}\!\big(\boldsymbol{\hat{x}}_n^{\left(j-1\right)}\big)\!\big)\!,
    \end{cases} \quad\forall n, j, t,
\end{equation}
which is the well-known soft-thresholding denoisers, but with a binary threshold depending on the estimation in the previous coherence block. This denoiser design does not depend on the distribution of $\boldsymbol{x}_n^{(j)}$'s, as shown in (\ref{sysm_si_st_denoiser}). Note that if the SI is not utilized, $\theta_{n,t}^{\left(j\right)}(f_{n,j}(\boldsymbol{\hat{x}}_n^{\left(j-1\right)}))=\theta_t^{(j)},\forall n,$ should hold over all the coherence blocks. In this case, the optimization of $\theta_t^{(j)}$ is considered in \cite{AMP,CAMP,AMP_NS}. However, under our considered SI-aided MMV-AMP framework, the challenge is how to design $\theta_{n,t}^{\left(j\right)}(f_{n,j}(\boldsymbol{\hat{x}}_n^{\left(j-1\right)}))$ in (\ref{iii_threshold}) based on the SI. In Section \ref{sec_unknown}, we will deal with this issue for SI-based binary threshold design under the soft-thresholding denoisers architecture.

On the other hand, if the distribution of $\boldsymbol{x}_{n,t}^{\left(j\right)}$'s , i.e., $\boldsymbol{\nu}_{n,M}$'s, is known in the linear model (\ref{model}), we can design the denoisers to minimize the MSE for the estimation of $\boldsymbol{x}_{n}^{\left(j\right)}$. Specifically, define a set of random vectors $\boldsymbol{\tilde{X}}_{n,t}^{\left(j\right)}\in\mathbb{C}^{M\times 1}$, $\boldsymbol{X}_n^{\left(j\right)}\in\mathbb{C}^{M\times 1}$, and $\boldsymbol{\hat{X}}_n^{\left(j-1\right)}\in\mathbb{C}^{M\times 1}$ which capture the distribution of $\boldsymbol{\tilde{x}}_{n,t}^{\left(j\right)}$, $\boldsymbol{x}_n^{\left(j\right)}$, and $\boldsymbol{\hat{x}}_n^{\left(j-1\right)}$, respectively, $\forall n,j$. The MMSE minimization problem is formulated as
\begin{align}
    \mathop{\text{minimize}}_{\eta_{n,t}^{\left(j\right)}\left ( \cdot, \diamond \right )}\mathop{\mathbb{E}}\left[\!\big \| \eta_{n, t}^{\left(j\right)} \!\big( \!\boldsymbol{\tilde{X}}_{n,t}^{\left(j\right)}, f_{n,j}\big(\boldsymbol{\hat{X}}_n^{\left(j-1\right)}\big)\!\big)\!-\! \boldsymbol{X}_{n}^{\left(j\right)}\big\|^2 \! \!\mid\! \boldsymbol{\tilde{X}}_{n,t}^{\left(j\right)} \!= \!\boldsymbol{\tilde{x}}_{n,t}^{\left(j\right)}, f_{n,j} \big( \boldsymbol{\hat{X}}_n^{\left(j-1\right)}\big) \!= \!\boldsymbol{\tilde{x}}_{n,\infty}^{\left(j-1\right)}\!\right], \quad \forall n,j,t.
\end{align}The optimal solution to the above problem leads to the MMSE denoisers:
\begin{align}\label{sec3_mmse}
    \eta_{n,t}^{\left(j\right)}\big(\boldsymbol{\tilde{x}}_{n,t}^{\left(j\right)}, f_{n,j}\big(\boldsymbol{\hat{x}}_n^{\left(j-1\right)}\big)\big)=\mathop{\mathbb{E}}\left[\boldsymbol{X}_n^{\left(j\right)}\mid \boldsymbol{\tilde{X}}_{n,t}^{\left(j\right)}=\boldsymbol{\tilde{x}}_{n,t}^{\left(j\right)}, f_{n,j}\big(\boldsymbol{\hat{X}}_n^{\left(j-1\right)}\big)=\boldsymbol{\tilde{x}}_{n,\infty}^{\left(j-1\right)}\right], \quad \forall n, j, t.
\end{align}
Note that if the SI in the previous coherence block is not utilized, the closed-form characterization of the MMSE denoisers is given in \cite{mMTC2} under the i.i.d. Rayleigh fading channel model. The challenge of the SI-based MMSE denoisers shown in (\ref{sec3_mmse}) lies in how to derive the conditional expectation if the SI is considered. In Section \ref{sec_known}, we will deal with this issue when the user channels $\boldsymbol{h}_n^{\left(j\right)}$'s follow the i.i.d. Rayleigh fading model.

\section{SI-based Denoiser Design with Unknown Channel Distribution}\label{sec_unknown}

In the case without knowledge of the channel distribution, the SI-based soft-thresholding denoisers are given in (\ref{sysm_si_st_denoiser}). The main technical issue is how to design the binary threshold $\theta_{n,t}^{\left(j\right)}(f_{n,j}(\boldsymbol{\hat{x}}_n^{\left(j-1\right)}))$'s in (\ref{iii_threshold}) based on the SI. Following \cite{AMP,CAMP,AMP_NS,AMP_PT}, in this section, we apply the minimax approach to design the threshold. Under this approach, we first obtain the least-favorable channel distribution $\boldsymbol{\mu}_{n,M}$ that leads to the maximum MSE for estimating $\boldsymbol{x}_n^{(j)}$ among all the channel distributions in $\mathcal{G}_{M}$, $\forall n$. Then, we design the binary threshold shown in (\ref{iii_threshold}) to minimize the above MSE. In other words, we try to design the binary threshold such that the worst-case performance of the soft-thresholding denoisers based MMV-AMP algorithm is optimized when the channel distribution is unknown.

In coherence block $j$, given the binary threshold $\theta_{1,n,t}^{(j)}$, $\theta_{2,n,t}^{(j)}$, and the channel distribution $\boldsymbol{\mu}_{n,M}$, we define the MSE for estimating $\boldsymbol{x}_n^{(j)}$ at the $t$-th iteration of the AMP algorithm with the SI-based soft-thresholding denoisers (\ref{sysm_si_st_denoiser}) as
\begin{align}
    &\text{MSE}_{n,t}^{(j)}(\theta_{1,n,t}^{(j)},\theta_{2,n,t}^{(j)},\boldsymbol{\mu}_{n,M})=\mathop{\mathbb{E}}_{_{\boldsymbol{X}_n^{(j)},\boldsymbol{\hat{X}}_n^{(j-1)}, \boldsymbol{V}_n^{(j)}}}\left[\big\|\eta_{n,t}^{(j)}\big ({\boldsymbol{X}}_{n}^{(j)}+(\boldsymbol{\Sigma}_t^{(j)})^\frac{1}{2}{\boldsymbol{V}}_n^{(j)},f_{n,j}( \boldsymbol{\hat{X}}_n^{(j-1)})\big) -\boldsymbol{X}_n^{\left(j\right)} \big\|^2\right]\notag\\
    =&Pr(\delta_n^{(j)}\!=\!0,\!\hat{\delta}_n^{(j-1)}\!=\!1)\!\!\mathop{\mathbb{E}}_{_{\boldsymbol{X}_n^{(j)},\boldsymbol{V}_n^{(j)}}}\!\!\left[\big\|\eta_{n,t}^{(j)}\!\big({\boldsymbol{X}}_{n}^{(j)}\!+\!(\boldsymbol{\Sigma}_t^{(j)})^\frac{1}{2}\boldsymbol{V}_n^{(j)}\!,\!f_{n,j}(\boldsymbol{\hat{X}}_n^{(j-1)}\!)\!\big) -\boldsymbol{X}_n^{(j)}\big\|^2 \mid\!\delta_n^{(j)}\!=\!0,\!\hat{\delta}_n^{(j-1)}\!=\!1\right]\notag\\
    +&Pr(\delta_n^{(j)}\!=\!0,\!\hat{\delta}_n^{(j-1)}\!=\!0)\!\!\mathop{\mathbb{E}}_{_{\boldsymbol{X}_n^{(j)},\boldsymbol{V}_n^{(j)}}}\!\!\left[\big\|\eta_{n,t}^{\left(j\right)}\big({\boldsymbol{X}_n}^{(j)}\!+\!(\boldsymbol{\Sigma}_t^{(j)})^\frac{1}{2}{\boldsymbol{V}}_n^{(j)}\!,\!f_{n,j}(\boldsymbol{\hat{X}}_n^{(j-1)}\!)\!\big) \!-\!\boldsymbol{X}_n^{\left(j\right)}\big\|^2 \mid\!\delta_n^{(j)}\!=\!0,\!\hat{\delta}_n^{(j-1)}\!=\!0\right]\notag\\
    +&Pr(\delta_n^{(j)}\!=\!1,\!\hat{\delta}_n^{(j-1)}\!=\!1)\!\!\!\!\!\mathop{\mathbb{E}}_{_{\boldsymbol{X}_n^{(j)},\boldsymbol{V}_n^{(j)}}}\!\!\left[\big\|\eta_{n,t}^{\left(j\right)}\big(\boldsymbol{X}_{n}^{(j)}\!+\!(\boldsymbol{\Sigma}_t^{(j)})^\frac{1}{2}\boldsymbol{V}_n^{(j)}\!,\!f_{n,j}(\boldsymbol{\hat{X}}_n^{(j-1)}\!)\!\big) \!-\!\boldsymbol{X}_n^{\left(j\right)}\big\|^2\! \mid\!\delta_n^{(j)}\!=\!1,\!\hat{\delta}_n^{(j-1)}\!=\!1\right]\notag\\
    +&Pr(\delta_n^{(j)}\!=\!1,\!\hat{\delta}_n^{(j-1)}\!=\!0)\!\!\!\!\! \mathop{\mathbb{E}}_{_{\boldsymbol{X}_n^{(j)}, \boldsymbol{V}_n^{(j)}}}\!\!\left[\big\|\eta_{n,t}^{(j)}\big(\boldsymbol{X}_{n}^{(j)}\!+\!(\boldsymbol{\Sigma}_t^{(j)})^\frac{1}{2}\boldsymbol{V}_n^{(j)}\!,\!f_{n,j}(\boldsymbol{\hat{X}}_n^{(j-1)}\!)\!\big)\!-\!\boldsymbol{X}_n^{(j)}\big\|^2\! \mid\!\delta_n^{(j)}\!=\!1,\!\hat{\delta}_n^{(j-1)}\!=\!0\right]\notag\\
    =&Pr(\delta_n^{(j)}\!=\!0,\!\hat{\delta}_n^{(j-1)}\!=\!1)R_{n,t}^{(j)}(\boldsymbol{0},\theta_{1,n,t}^{(j)})
    +Pr(\delta_n^{(j)}\!=\!1,\!\hat{\delta}_n^{(j-1)}\!=\!1)\mathop{\mathbb{E}}_{_{\boldsymbol{h}_n^{(j)}\!\sim\!\boldsymbol{\mu}_{n,M}}}\!\left[R_{n,t}^{(j)}(\boldsymbol{h}_n^{(j)},\theta_{1,n,t}^{(j)})\right]\notag\\
    +&Pr(\!\delta_n^{(j)}\!=\!0,\!\hat{\delta}_n^{(j-1)}\!=\!0)R_{n,t}^{(j)}(\boldsymbol{0},\theta_{2,n,t}^{(j)})
    \!+\!Pr(\delta_n^{(j)}\!=\!1,\!\hat{\delta}_n^{(j-1)}\!=\!0)\!\!\!\mathop{\mathbb{E}}_{_{\boldsymbol{h}_n^{(j)}\!\sim\!\boldsymbol{\mu}_{n,M}}}\!\!\!\!\!\left[R_{n,t}^{(j)}(\boldsymbol{h}_n^{(j)},\theta_{2,n,t}^{(j)})\right],\quad\forall n, j ,t,\label{iv_mse_2}
\end{align}
where
\begin{align}\label{def_risk}
   R_{n,t}^{(j)}({\boldsymbol{x}}_{n}^{(j)},\theta_{i,n,t}^{(j)})=\mathop{\mathbb{E}}_{_{\boldsymbol{V}_n^{(j)}}}\!\!\left[\big\|g_{n,t}^{\left(j\right)}\big(\boldsymbol{x}_n^{(j)}+(\boldsymbol{\Sigma}_t^{(j)})^\frac{1}{2}\boldsymbol{V}_n^{(j)},\theta_{i,n,t}^{(j)}\big) \!-\!\boldsymbol{x}_n^{(j)}\big\|^2\right],\quad i=1,2,\forall n,j,t,
\end{align}and
\begin{align}\label{def_g}
    g_{n,t}^{(j)}(\tilde{\boldsymbol{x}}_{n,t}^{(j)},\theta_{i,n,t}^{(j)})=(\tilde{\boldsymbol{x}}_{n,t}^{(j)}-\theta_{i,n,t}^{(j)}\frac{\tilde{\boldsymbol{x}}_{n,t}^{(j)}}{\|\tilde{\boldsymbol{x}}_{n,t}^{(j)}\|})\mathop{\mathbb{I}}(\|\tilde{\boldsymbol{x}}_{n,t}^{(j)}\|\ge\theta_{i,n,t}^{(j)}), \quad i=1,2,\forall n,j,t,
\end{align}with $\mathop{\mathbb{I}}(\cdot)$ denoting the indicator function. Then, the optimal threshold for the least-favorable distribution can be obtained by solving the following problem
\begin{align}\label{minimax}
    (\theta_{1,n,t}^{(j),\ast}, \theta_{2,n,t}^{(j),\ast})=\underset{\theta_{1,n,t}^{(j)}, \theta_{2,n, t}^{(j)}}{\operatorname{argmin}} \underset{\boldsymbol{\mu}_{n,M}}{\operatorname{max}}~\mathrm{MSE}_{n,t}^{(j)}(\theta_{1,n,t}^{(j)}, \theta_{2,n,t}^{(j)},\boldsymbol{\mu}_{n,M}),\quad \forall n,j,t.
\end{align}Note that in (\ref{iv_mse_2}), both the probabilities $Pr(\delta_n^{(j)},\!\hat{\delta}_n^{(j-1)})$'s and the channel estimation MSEs $\mathop{\mathbb{E}}_{_{\boldsymbol{h}_n^{(j)}\sim\boldsymbol{\mu}_{n,M}}} \big[R_{n,t}^{(j)}(\boldsymbol{h}_n^{(j)},\theta_{i,n,t}^{(j)})\big]$'s, $i=1,2$, are functions of the distribution of $\boldsymbol{h}_n^{(j)}$'s, i.e., $\boldsymbol{\mu}_{n,M}$'s. Especially, the device activity detection is based on (\ref{llr_detection}), and the distribution $\boldsymbol{\mu}_{n,M}$ will thus affect the probabilities of the events that $\hat{\delta}_n^{(j-1)}=1$ and $\hat{\delta}_n^{(j-1)}=0$, $\forall n$. As a result, even given the binary threshold, it is non-trivial to find the least-favorable distribution to maximize the MSE in (\ref{iv_mse_2}). This motivates us to find an approximate least-favorable channel distribution that has the closed-form probability density function (PDF) and can approach the MSE achieved by the least-favorable channel distribution.

It was shown in \cite{mMTC2} that under the conventional AMP algorithm without utilizing SI, the performance of the device activity detection is very good. Especially, when the number of antennas goes to infinity, perfect device activity detection can be achieved, i.e., $\hat{\delta}_n^{(j-1)}=\delta_n^{(j-1)}$ holds almost surely for each device $n$. As a result, in practice, we usually have $Pr(\delta_n^{(j)},\hat{\delta}_n^{(j-1)})\approx Pr(\delta_n^{(j)},{\delta}_n^{(j-1)})$, $\forall n$. In other words, the channel distribution does not affect $Pr(\delta_n^{(j)},\hat{\delta}_n^{(j-1)})$ too much. However, due to the use of the non-orthogonal pilot sequences, the channel estimation MSEs are large even under the perfect detection case\cite{mMTC8}. As a result, the channel distribution has a significant impact on channel estimation MSEs. Therefore, instead of focusing on the exact least-favorable distribution that can maximize the MSE in (\ref{iv_mse_2}), which is complicated, in the following, we focus on the PDF of the distribution that tends to maximize the channel estimation MSEs $\mathop{\mathbb{E}}_{_{\boldsymbol{h}_n^{(j)}\sim\boldsymbol{\mu}_{n,M}}}[ R_{n,t}^{(j)}(\boldsymbol{h}_n^{(j)},\theta_{i,n,t}^{(j)})]$'s, $i=1,2$. It was shown in \cite{AMP,CAMP} that in the SMV case, i.e., $M=1$ and $h_n^{(j)}$ is a scalar, $\forall n,j$, $\mathop{\mathbb{E}}_{_{{h}_n^{(j)}\sim{\mu}_{n,1}}}\big[R_{n,t}^{(j)}({h}_n^{(j)},\theta_t^{(j)})\big]$ is maximized when $|{h}_n^{(j)}|$ goes to infinity, i.e., ${\mu}_{n,1}=\delta_{\infty}(|{h}_{n}^{(j)}|)$, $\forall n, j$, where $\delta_{\infty}(\cdot)$ denotes the point mass at $\infty$. In the case when $M>1$ such that $\boldsymbol{h}_n^{(j)}$ is a vector, $\forall n,j$, \cite{AMP_PT} showed that the maximum of $\mathop{\mathbb{E}}_{_{\boldsymbol{h}_n^{(j)}\sim{\boldsymbol{\mu}}_{n,M}}}\big[R_{n,t}^{(j)}(\boldsymbol{h}_n^{(j)},\theta_t^{(j)})\big]$ is achieved when $\|\boldsymbol{h}_n^{(j)}\|$ goes to infinity, $\forall n,j$. Since we assume in this paper that the user channels are i.i.d. over $m$, we define the following channel distribution as the approximate least-favorable channel distribution to maximize the MSE given in (\ref{iv_mse_2})
\begin{align}\label{each_least_favorable}
    \boldsymbol{\hat{\mu}}_{n,M}^{\ast}={\hat{\mu}}_{n}^{\ast}\times\cdots\times{\hat{\mu}}_{n}^{\ast}=\delta_{\infty}(|{h}_{n,1}^{(j)}|)\times\cdots\times\delta_{\infty}(|{h}_{n,M}^{(j)}|), \quad\forall n, j.
\end{align}

With the approximate least-favorable channel distribution given in (\ref{each_least_favorable}), we simplify the MSE in (\ref{iv_mse_2}). First, we have the following lemma.

\begin{lemma}\label{iv_lemma}
Consider the SI-based soft-thresholding denoisers given in (\ref{sysm_si_st_denoiser}) and the approximate least-favorable channel distribution given in (\ref{each_least_favorable}). Under the asymptotic regime where $N, K, L\to \infty$ with fixed $N/L$ and $N/K$, the state evolution defined by (\ref{state_evolution}) always stays as a diagonal matrix with identical diagonal entries, i.e.,
\begin{align}\label{dia}
    \boldsymbol{\Sigma}_t^{\left(j\right)} = \big(\tau_t^{\left(j\right)}\big)^2\boldsymbol{I}, ~~~ \forall j,t.
\end{align}
\end{lemma}
\begin{IEEEproof}
Lemma \ref{iv_lemma} can be proved using the similar induction method as that in Appendix B of \cite{mMTC2}, which is omitted here due to the space limitation.
\end{IEEEproof}
With the simplified state shown in (\ref{dia}), we are able to characterize the MSE shown in (\ref{iv_mse_2}) under the approximate least-favorable channel distribution (\ref{each_least_favorable}).
\begin{theorem}\label{theorem1}
Consider the SI-based soft-thresholding denoisers given in (\ref{sysm_si_st_denoiser}), the approximate least-favorable distribution given in (\ref{each_least_favorable}), and the temporal correlation model for device activity shown in (\ref{p_transition}), under the asymptotic regime where $N, K, L\to \infty$ with fixed $N/L$ and $N/K$. Define
\begin{align}
    \varpi^{(j)}\!(\theta_{i,n,t}^{(j)})\!&=\!\frac{({\theta_{i,n,t}^{\left(j\right)}})^2\overline{\Gamma}\big(M\!,\!(\frac{\theta_{i,n,t}^{\left(j\right)}}{\tau_t^{(j)}})^2\big)\!-\!2{\theta_{i,n,t}^{\left(j\right)}}{\tau_t^{(j)}}\overline{\Gamma}\big(M\!+\!\frac{1}{2},(\frac{\theta_{i,n,t}^{\left(j\right)}}{\tau_t^{(j)}})^2\big)\!+\!(\tau_t^{(j)})^2\overline{\Gamma}\big(\!M\!+\!1,\!(\frac{\theta_{i,n,t}^{\left(j\right)}}{\tau_t^{(j)}})^2\big)}{\Gamma(M)},\notag\\ &\quad\quad\quad\quad\quad\quad\quad\quad\quad\quad\quad\quad\quad\quad\quad\quad\quad\quad\quad\quad\quad\quad\quad\quad\quad i=1,2, \forall n, j, t,
\end{align}and
\begin{align}\label{Th1_P_fa}
    \varsigma_{\infty}^{(j-1)}=\frac{\overline{\Gamma}(M,(l^{(j-1)})^{2}(\tau_{\infty}^{(j-1)})^{-2})}{\Gamma(M)}, \quad\forall j,
\end{align}
where $\Gamma\left(\cdot\right)$ and $\overline{\Gamma}\left(\cdot,\cdot\right)$ denote the Gamma function and the upper incomplete Gamma function \cite{gamma}, respectively. Under the approximate least-favorable distribution given in (\ref{each_least_favorable}), the MSE given in (\ref{iv_mse_2}) is expressed as
\begin{align}\label{th1_worstcasemse}
    {\text{MSE}}_{n,t}^{(j)}(\theta_{1,n,t}^{(j)},\theta_{2,n,t}^{(j)},\boldsymbol{\hat{\mu}}_{n,M}^{\ast})&={\text{MSE}}_{1,n,t}^{(j)}(\theta_{1,n,t}^{(j)},\boldsymbol{\hat{\mu}}_{n,M}^{\ast})+{\text{MSE}}_{2,n,t}^{(j)}(\theta_{2,n,t}^{(j)},\boldsymbol{\hat{\mu}}_{n,M}^{\ast}), \quad\forall n,j,t,
\end{align}where
\begin{align}
    {\text{MSE}}_{1,n,t}^{(j)}(\theta_{1,n,t}^{(j)},\boldsymbol{\hat{\mu}}_{n,M}^{\ast})&=\big(\beta(1-\lambda)\varsigma_{\infty}^{(j-1)}+\alpha\lambda\big)\big((\tau_t^{(j)})^{2}M+({\theta_{1,n,t}^{\left(j\right)}})^2\big)\notag\\&+\big((1-\beta)(1-\lambda)\varsigma_{\infty}^{(j-1)}+(1-\alpha)\lambda\big)\varpi^{(j)}(\theta_{1,n,t}^{(j)}), \quad\forall n,j,t,
\end{align}and
\begin{align}
    {\text{MSE}}_{2,n,t}^{(j)}(\theta_{2,n,t}^{(j)},\boldsymbol{\hat{\mu}}_{n,M}^{\ast})&=\beta(1-\lambda)(1-\varsigma_{\infty}^{(j-1)})\big((\tau_t^{(j)})^{2}\!M+\!({\theta_{2,n,t}^{\left(j\right)}})^2\big)\notag\\&+(1-\beta)(1-\lambda)(1-\varsigma_{\infty}^{(j-1)})\varpi^{(j)}(\theta_{2,n,t}^{(j)}),\quad \forall n,j,t.
\end{align}
\end{theorem}
\begin{IEEEproof}
Please refer to Appendix \ref{worstcasemse}.
\end{IEEEproof}

After characterizing the MSE under the approximate least-favorable channel distribution (\ref{each_least_favorable}) in Theorem \ref{theorem1}, we aim to find the best binary threshold to minimize the MSE in (\ref{th1_worstcasemse}):
\begin{align}\label{pro_optimalthresholds}
    \hat{\theta}_{i,n,t}^{(j),\ast}=\underset{\theta_{i,n,t}^{(j)}}{\operatorname{argmin}} ~\mathrm{MSE}_{i,n,t}^{(j)}(\theta_{i,n,t}^{(j)}, \boldsymbol{\hat{\mu}}_{n,M}^{\ast}),\quad i=1,2, \forall n,j,t.
\end{align}The optimal solution to problem (\ref{pro_optimalthresholds}) can be found via Theorem \ref{theorem2}.
\begin{theorem}\label{theorem2}
Define two functions of ${\theta}_{1,n,t}^{(j)}$ and ${\theta}_{2,n,t}^{(j)}$ as follows
\begin{align}\label{free_par1}
    &f_1({\theta}_{1,n,t}^{(j)})\!=\![(1\!-\!\beta)(1\!-\!\lambda)\varsigma_{\infty}^{(j-1)}\!+\!(1\!-\!\alpha)\lambda]\xi^{\left(j\right)}({\theta}_{1,n,t}^{(j)})\!+\![\beta(1\!-\!\lambda)\varsigma_{\infty}^{(j-1)}\!+\!\alpha\lambda]{\theta}_{1,n,t}^{(j)},\quad\forall n, j, t,\\
    &f_2({\theta}_{2,n,t}^{(j)})\!=\!(1-\beta)(1-\lambda)(1-\varsigma_{\infty}^{(j-1)})\xi^{\left(j\right)}({\theta}_{2,n,t}^{(j)})\!+\!\beta(1-\lambda)(1-\varsigma_{\infty}^{(j-1)}){\theta}_{2,n,t}^{(j)},\quad\forall n, j, t,\label{free_par2}
\end{align}
with $\varsigma_{\infty}^{(j-1)}$ given in (\ref{Th1_P_fa}) and
\begin{equation}
   \xi^{\left(j\right)}({\theta}_{i,n,t}^{(j)})= \frac{{\theta}_{i,n,t}^{(j)}\overline{\Gamma}\big(M,(\frac{{\theta}_{i,n,t}^{(j)}}{\tau_t^{(j)}})^2\big)-\tau_t^{(j)}\overline{\Gamma}\big(M+\frac{1}{2},(\frac{{\theta}_{i,n,t}^{(j)}}{\tau_t^{(j)}})^2\big)}{\Gamma(M)},\quad i=1,2, \forall n,j,t.
\end{equation}There always exists a unique positive solution $\hat{\theta}_{1,n,t}^{(j),\ast}$ to the equation $f_1({\theta}_{1,n,t}^{(j)})=0$ and a unique positive solution $\hat{\theta}_{2,n,t}^{(j),\ast}$ to the equation $f_2({\theta}_{2,n,t}^{(j)})=0$. Then, $\hat{\theta}_{i,n,t}^{(j),\ast}, i=1,2, \forall n,j,t$, is the optimal solution to problem (\ref{pro_optimalthresholds}).
\end{theorem}
\begin{IEEEproof}
Please refer to Appendix \ref{threshold}.
\end{IEEEproof}

Note that in Theorem \ref{theorem2}, the solution to (\ref{free_par1}) and (\ref{free_par2}) can be easily obtained by the bisection method, since both $f_1({\theta}_{1,n,t}^{(j)})$ and $f_2({\theta}_{2,n,t}^{(j)})$ are increasing functions over ${\theta}_{1,n,t}^{(j)}$ and ${\theta}_{2,n,t}^{(j)}$, $\forall n,j,t$. As a result, we may take $\hat{\theta}_{i,n,t}^{(j),\ast}, i=1,2, \forall n,j,t$, as the binary threshold to the SI-based soft-thresholding denoisers (\ref{sysm_si_st_denoiser}) in the SI-aided MMV-AMP algorithm when the channel distribution is unknown. Since the binary threshold is obtained based on the approximate least-favorable channel distribution, it is anticipated that the performance under other channel distributions will be further improved in general.

\begin{figure}[t]
	\centering
	\includegraphics[height=6.6cm]{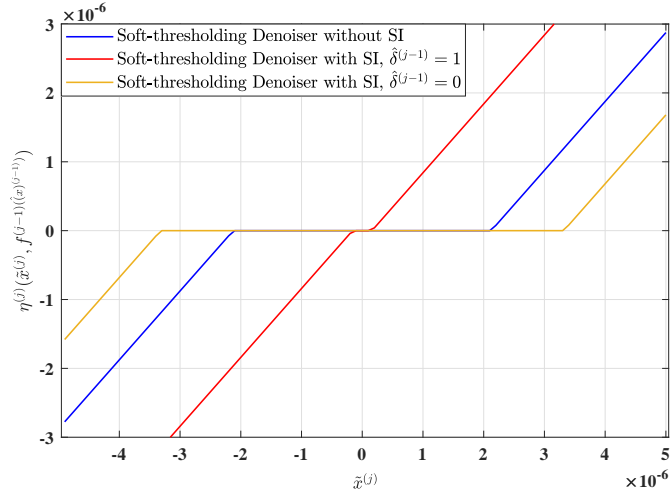}
	\vspace{-5mm}
	\caption{Comparison of soft-thresholding denoisers with or without using SI.}
	\label{Fig_st_denoiser}
	\vspace{-10mm}
\end{figure}

In the following, we provide a numerical example to show the effect of the SI on the SI-based soft-thresholding denoiser design when the device activity is temporally correlated. In this example, we set $M=1$, $\lambda=0.1$, $\alpha=0.91>\lambda$, $\beta=0.01<\lambda$, $\varsigma_\infty^{\left(j-1\right)}=0$ and $\tau^{\left(j\right)}=2\times10^{-6}$. Fig. \ref{Fig_st_denoiser} shows the SI-based soft-thresholding denoisers when a device was detected as an active and inactive device in the previous coherence block, respectively, as well as the denoisers without using SI. Compared to the denoisers without using SI, it can be observed that when a device is detected to be active (or inactive) previously, the SI-based soft-thresholding denoisers tend to detect this device as an active (or inactive) device in the current coherence block.

\section{SI-based Denoiser Design with known Channel Distribution}\label{sec_known}

In this section, we consider the case where the PDF of the channel is unknown, and introduce how to adopt the Bayesian approach to design the SI-based MMSE denoisers $\eta_{n,t}^{\left(j\right)}\left ( \cdot, \diamond \right )$'s for signal recovery. Specifically, we consider the Rayleigh fading channel model as an example, i.e., $\boldsymbol{h}_n^{(j)}\sim \mathcal{CN}(\boldsymbol{0},\gamma_n\boldsymbol{I})$, $\forall n,j$, where $\gamma_n$ is the path loss of device $n$. It is worth noting that the SI-based MMSE denoiser design can be extended to other channel models as long as the channel distribution is known.

At the $(t+1)$-th iteration of the AMP algorithm in coherence block $j$, the available information includes $ \boldsymbol{x}_{n,t}^{\left(j\right)}+(\boldsymbol{R}_t^{\left(j\right)})^H\boldsymbol{s}_{n}$ from the current coherence block whose distribution is modeled by (\ref{model}) and the SI from the previous coherence block $ \boldsymbol{\hat{x}}_n^{\left(j-1\right)}+(\boldsymbol{R}_\infty^{\left(\!j-1\!\right)})^H \boldsymbol{s}_{n}$ whose distribution is modeled by (\ref{state_estimation}). Based on the above information, the SI-based MMSE denoisers can be further expressed as
\begin{align}\label{expformmse}
    \mathop{\mathbb{E}}[&\boldsymbol{X}_n^{\left(j\right)}\mid\boldsymbol{X}_n^{\left(j\right)}+\big(\boldsymbol{\Sigma}_t^{\left(j\right)}\big )^\frac{1}{2}\boldsymbol{V}_n^{(j)} = \boldsymbol{\tilde{x}}_{n,t}^{\left(j\right)},\boldsymbol{{X}}_n^{\left(j-1\right)}+\big(\boldsymbol{\Sigma}_\infty^{\left(j-1\right)}\big )^\frac{1}{2}\boldsymbol{V}_n^{(j-1)}=\boldsymbol{\tilde{x}}_{n,\infty}^{\left(j-1\right)}], \quad\forall n,j,t.
\end{align}
To characterize the SI-based MMSE denoisers (\ref{expformmse}), we first derive the following result.
\begin{lemma}\label{lemma_appc}
Let $\boldsymbol{\tilde X}_{n,t}^{\left(j\right)}=\boldsymbol{X}_{n}^{\left(j\right)}+(\boldsymbol{\Sigma}_t^{\left(j\right)})^\frac{1}{2}\boldsymbol{V}_n^{(j)}$ and $\boldsymbol{\tilde X}_{n,\infty}^{\left(j-1\right)}=\boldsymbol{X}_n^{\left(j-1\right)}+(\boldsymbol{\Sigma}_\infty^{\left(j-1\right)})^\frac{1}{2}\boldsymbol{V}_n^{(j-1)}$ where $\boldsymbol{X}_n^{\left(j\right)}$, $\boldsymbol{X}_n^{\left(j-1\right)}$, $\boldsymbol{V}_n^{(j)}$, and $\boldsymbol{V}_n^{(j-1)}$ capture the distribution of  $\boldsymbol{x}_n^{\left(j\right)}$, $\boldsymbol{x}_n^{\left(j-1\right)}$, $\boldsymbol{v}_n^{(j)}$, and $\boldsymbol{v}_n^{(j-1)}$, respectively. Define
\begin{align}
    \phi\big(\boldsymbol{\tilde x}_{n,t}^{\left(j\right)},\boldsymbol{\tilde x}_{n,\infty}^{\left(j-1\right)}\big)&=\frac{1}{1+\frac{1-\lambda}{\lambda}\frac{\beta \boldsymbol{\psi}_{\boldsymbol{\Sigma}_t^{\left(j\right)}}(\boldsymbol{\tilde x}_{n,t}^{\left(j\right)})\boldsymbol{\psi}_{\gamma_n\boldsymbol{I}+\boldsymbol{\Sigma}_\infty^{\left(j-1\right)}}(\boldsymbol{\tilde{x}}_{n,\infty}^{\left(j-1\right)})+(1-\beta) \boldsymbol{\psi}_{\boldsymbol{\Sigma }_t^{\left(j\right)}}(\boldsymbol{\tilde x}_{n,t}^{\left(j\right)})\boldsymbol{\psi}_{\boldsymbol{\Sigma} _\infty^{\left(j-1\right)}}(\boldsymbol{\tilde{x}}_{n,\infty}^{\left(j-1\right)})}{\alpha\boldsymbol{\psi}_{\gamma_n\boldsymbol{I}+\boldsymbol{\Sigma }_t^{\left(j\right)}}(\boldsymbol{\tilde x}_{n,t}^{\left(j\right)})\boldsymbol{\psi}_{\gamma_n\boldsymbol{I}+\boldsymbol{\Sigma}_\infty^{\left(j-1\right)}}(\boldsymbol{\tilde{x}}_{n,\infty}^{\left(j-1\right)})+(1-\alpha)\boldsymbol{\psi}_{\gamma_n\boldsymbol{I}+\boldsymbol{\Sigma}_t^{\left(j\right)}}(\boldsymbol{\tilde x}_{n,t}^{\left(j\right)})\boldsymbol{\psi}_{\boldsymbol{\Sigma} _\infty^{\left(j-1\right)}}(\boldsymbol{\tilde{x}}_{n,\infty}^{\left(j-1\right)})}},\notag\\
    &\qquad\qquad\qquad\qquad\qquad\qquad\qquad\qquad\qquad\qquad\qquad\qquad\qquad\quad\forall n,j,t.
\end{align}
Then,
\begin{align}\label{comlicated_mmse}
    \mathop{\mathbb{E}}[\boldsymbol{X}_n^{\left(j\right)}\mid \boldsymbol{\tilde{X}}_{n,t}^{\left(j\right)}=\boldsymbol{\tilde x}_{n,t}^{\left(j\right)}, \boldsymbol{\tilde{X}}_{n,\infty}^{\left(j-1\right)}=\boldsymbol{\tilde x}_{n,\infty}^{\left(j-1\right)}]= \phi(\boldsymbol{\tilde x}_{n,t}^{\left(j\right)},\boldsymbol{\tilde x}_{n,\infty}^{\left(j-1\right)}){\gamma_n}({\gamma_n\boldsymbol{I}+\boldsymbol{\Sigma}_t^{\left(j\right)}})^{-1}{\boldsymbol{\tilde x}_{n,t}^{\left(j\right)}}, \quad\forall n,j,t,
\end{align}
and
\begin{align}
    &\mathop{\mathbb{E}}[\boldsymbol{X}_n^{\left(j\right)}{(\boldsymbol{X}_n^{\left(j\right)})}^{H}\mid \boldsymbol{\tilde{X}}_{n,t}^{\left(j\right)}=\boldsymbol{\tilde x}_{n,t}^{\left(j\right)}, \boldsymbol{\tilde{X}}_{n,\infty}^{\left(j-1\right)}=\boldsymbol{\tilde x}_{n,\infty}^{\left(j-1\right)}] = \phi(\boldsymbol{\tilde x}_{n,t}^{\left(j\right)},\boldsymbol{\tilde x}_{n,\infty}^{\left(j-1\right)})\left({\gamma_n}\boldsymbol{I}-\gamma_n^2(\gamma_n\boldsymbol{I}+\boldsymbol{\Sigma}_t^{\left(j\right)})^{-1}\right.\notag\\
    &\left.~~~~~~~~~~~~~~~~~~~~~~~~~~~~~~~~~~~+\gamma_n^2(\gamma_n\boldsymbol{I}+\boldsymbol{\Sigma}_t^{\left(j\right)})^{-1}\boldsymbol{\tilde x}_{n,t}^{\left(j\right)}(\boldsymbol{\tilde x}_{n,t}^{\left(j\right)})^H(\gamma_n\boldsymbol{I}+\boldsymbol{\Sigma}_t^{\left(j\right)})^{-1}\right),\quad\forall n,j,t.
\end{align}
\end{lemma}
\begin{IEEEproof}
Please refer to Appendix \ref{app_mmse_derivation}.
\end{IEEEproof}
With Lemma \ref{lemma_appc}, the state evolution can be further simplified.

\begin{lemma}\label{dia_state_evo}
Given the SI-based MMSE denoisers in (\ref{expformmse}), under the asymptotic regime where $N,L,K\to \infty$ with fixed $N/L$ and $N/K$. The matrix $\boldsymbol{\Sigma}_t^{\left(j\right)}$ generated by the state evolution (\ref{state_evolution}) always stays as a diagonal matrix with identical diagonal entries, i.e.,
\begin{equation}\label{lemmav_dia}
    \boldsymbol{\Sigma}_t^{\left(j\right)} = \big(\tau_t^{\left(j\right)}\big)^2\boldsymbol{I}, \quad \forall j,t.
\end{equation}
\end{lemma}
\begin{IEEEproof}
Lemma \ref{dia_state_evo} can be proved using the similar induction method as that in Appendix B of \cite{mMTC2}, which is omitted here due to space limitation.
\end{IEEEproof}
With the simplified state shown in Lemma \ref{dia_state_evo}, we can thus derive the simplified SI-based MMSE denoisers in Theorem \ref{theorem3}.

\begin{theorem}\label{theorem3}
Consider the SI-aided MMV-AMP algorithm given by (\ref{amp_x}) and (\ref{amp_r}) under the temporal correlation model for device activity shown in (\ref{p_transition}). Define
\begin{align}
    \Delta_{n,t}^{\left(j\right)}=\big(\tau_{t}^{\left(j\right)}\big)^{-2}-\big(\big(\tau_t^{\left(j\right)}\big)^{2}+\gamma_n\big)^{-1}, \quad \forall n,j,t. \label{Delta1}
\end{align}
Under the asymptotic regime where $N,L, K\to \infty$ with fixed $N/L$ and $N/K$, the MMSE denoisers (\ref{expformmse}) in coherence block $j$ with the SI given in (\ref{SI}) are expressed as:
\begin{equation}\label{simplified denoiser}
    \eta_{n,t}^{\left(j\right)} \big(\boldsymbol{\tilde{x}}_{n,t}^{\left(j\right)}, f_{n,j}\big(\boldsymbol{\hat{x}}_n^{\left(j-1\right)}\big)\big)
    =\frac{\gamma_n\big(\gamma_n+\big(\tau_t^{\left(j\right)}\big)^2\big)^{-1}\boldsymbol{\tilde{x}}_{n,t}^{\left(j\right)}}{1+\frac{1-\lambda}{\lambda}\mu_{n,t}^{\left(j\right)}\times \frac{\beta+(1-\beta)\mu_{n,\infty}^{\left(j-1\right)} }{\alpha+(1-\alpha)\mu_{n,\infty}^{\left(j-1\right)}}}, \quad\forall n,j,t,
\end{equation}
where
\begin{equation}\label{mu}
    \mu_{n,t}^{\left(j\right)}= \left(\frac{\big(\tau_t^{\left(j\right)}\big)^2+\gamma_n}{\big(\tau_t^{\left(j\right)}\big)^2}\right)^M{\rm exp}\big(-\Delta_{n,t}^{\left(j\right)}\|\boldsymbol{\tilde{x}}_{n,t}^{\left(j\right)}\|^2\big),\quad\forall n,j,t,
\end{equation}and $(\tau_{\infty}^{(j-1)})^2$ can be obtained from the state evolution (\ref{state_evolution}) and (\ref{lemmav_dia}) after AMP converges in coherence block $j-1$.
\end{theorem}
\begin{IEEEproof}
	By plugging (\ref{lemmav_dia}) into (\ref{comlicated_mmse}), the results in Theorem \ref{theorem3} can be readily derived.
\end{IEEEproof}

To gain insights from Theorem \ref{theorem3}, we discuss several special cases. First, if the user activity is independent over different coherence blocks, i.e., $\alpha=\beta=\lambda$ such that $Pr(\delta_n^{\left(j\right)}\mid \delta_n^{\left(j-1\right)}) =Pr(\delta_n^{\left(j\right)})$, $\forall n,j$, the denoisers in (\ref{simplified denoiser}) will reduce to
\begin{equation}\label{denoiserwithoutsi}
    \eta_{n,t}^{\left(j\right)} \!\big(\!\boldsymbol{\tilde{x}}_{n,t}^{\left(j\right)}, f_{n,j}(\boldsymbol{\hat{x}}_n^{\left(j-1\right)})\!\big)\!
    \!=\!\frac{\gamma_n\!\left(\!\gamma_n\!+\!(\tau_t^{(j)})^2\!\right)^{-1}\!\boldsymbol{\tilde{x}}_{n,t}^{\left(j\right)}}{1+\frac{1-\lambda}{\lambda}\mu_{n,t}^{\left(j\right)}},\quad \forall n,j,t,
\end{equation}
which are the MMSE denoisers proposed in \cite{mMTC2, mMTC3} without taking SI into account. This is because if there is no temporal correlation in the user activity, SI will have no effect on the MMSE denoiser design. Second, if $(\tau_{\infty}^{(j-1)})^2\to\infty$, it can be shown from (\ref{mu}) that $\mu_{n,\infty}^{\left(j-1\right)}=1$, $\forall n$. Then, the MMSE denoisers shown in (\ref{simplified denoiser}) will also reduce to the MMSE denoisers in (\ref{denoiserwithoutsi}) proposed in \cite{mMTC2, mMTC3} without taking SI into account. This is because according to (\ref{state_estimation}) and (\ref{dia}), $(\tau_{\infty}^{(j-1)})^2$ can be viewed as the equivalent noise power for estimating $\boldsymbol{X}^{(j-1)}$ by AMP. If this noise power is infinite but the power of each row in $\boldsymbol{X}^{(j-1)}$ is finite, then the estimation does not provide any useful information for the estimation in the next block, despite the existence of temporal correlation in activity.
\begin{figure}[t]
	\centering
	\includegraphics[height=6.6cm]{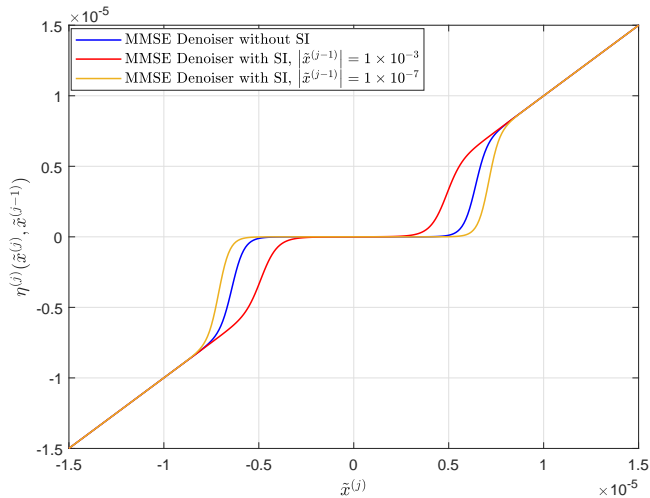}
	\vspace{-5mm}
    \caption{Comparison of MMSE denoisers with or without using SI.} \label{Fig_mmsedenoiser}
	\vspace{-8mm}
\end{figure}

Next, we provide a numerical example to show the gain of using SI in the MMSE denoiser design when the device activity is temporally correlated. In this example, we set $M=1$, $\lambda=0.1$, $\alpha=0.91>\lambda$, $\beta=0.01<\lambda$, $\tau^{\left(j\right)}=\tau^{\left(j-1\right)}=2\times10^{-6}$, and $\gamma=1\times10^{-8}$. Fig. \ref{Fig_mmsedenoiser} shows the SI-based MMSE denoisers when $|\tilde{x}^{\left(j-1\right)}|=1\times10^{-3}$ and $|\tilde{x}^{\left(j-1\right)}|=1\times10^{-7}$ as well as the denoisers without using SI. Compared to the denoisers without using SI, it is observed that when $|\tilde{x}^{\left(j-1\right)}|$ is larger (or smaller), i.e., the device tends to be detected as an active (or inactive) device in the previous block, the SI-based MMSE denoisers estimates $x^{(j)}$ as zero over a smaller (or larger) range of $\tilde{x}^{(j)}$, i.e., the device tends to be detected as an active (inactive) device in the current coherence block.
\vspace{-3mm}
\section{Numerical Results}\label{sec_num}

In this section, we provide numerical results to evaluate the performance of the proposed SI-aided MMV-AMP framework in massive IoT connectivity systems without or with channel distribution information. We assume that there are $N=2000$ devices randomly located in a cell of radius $R=500$ meters (m). The channels are generated according to the Rayleigh fading model where the path loss is modeled as $-128.1-36.7\log_{10}(d_n)$ in dB, with $d_n$ in kilometers (km) denoting the distance from device $n$ to the BS. We consider the communication over $J=10$ coherence blocks, while in each coherence block, we have  $\lambda=0.1$, $\alpha=0.55$, and $\beta=0.05$. Next, the user transmit power is set as $23$ dBm. Last, the power spectrum density of the noise is $-169$ dBm/Hz, while the bandwidth of the channel is set as $10$ MHz.

\subsection{Activity Detection and Channel Estimation Performance with Unknown Channel Distribution}\label{Sec_A}

First, we consider the case that the BS does not know the channel distribution and evaluate the performance of device activity detection and channel estimation achieved by our proposed SI-aided MMV-AMP frameowrk with soft-thresholding denoisers. We consider two benchmark schemes. Specifically, to obtain a performance upper bound, we consider the scheme where the SI is always perfect, i.e., $\hat{\delta}_n^{(j-1)}=\delta_{n}^{(j-1)}$, $\forall n,j$, and the optimal binary threshold in (\ref{iii_threshold}) is derived using \emph{exhaustive search}; while to obtain a performance lower bound, we consider the scheme where the SI is not utilized in the AMP algorithm \cite{CAMP}.

\begin{figure}[t]
	\centering
	\subfigure[Activity Detection Performance]{\includegraphics[height=6cm]{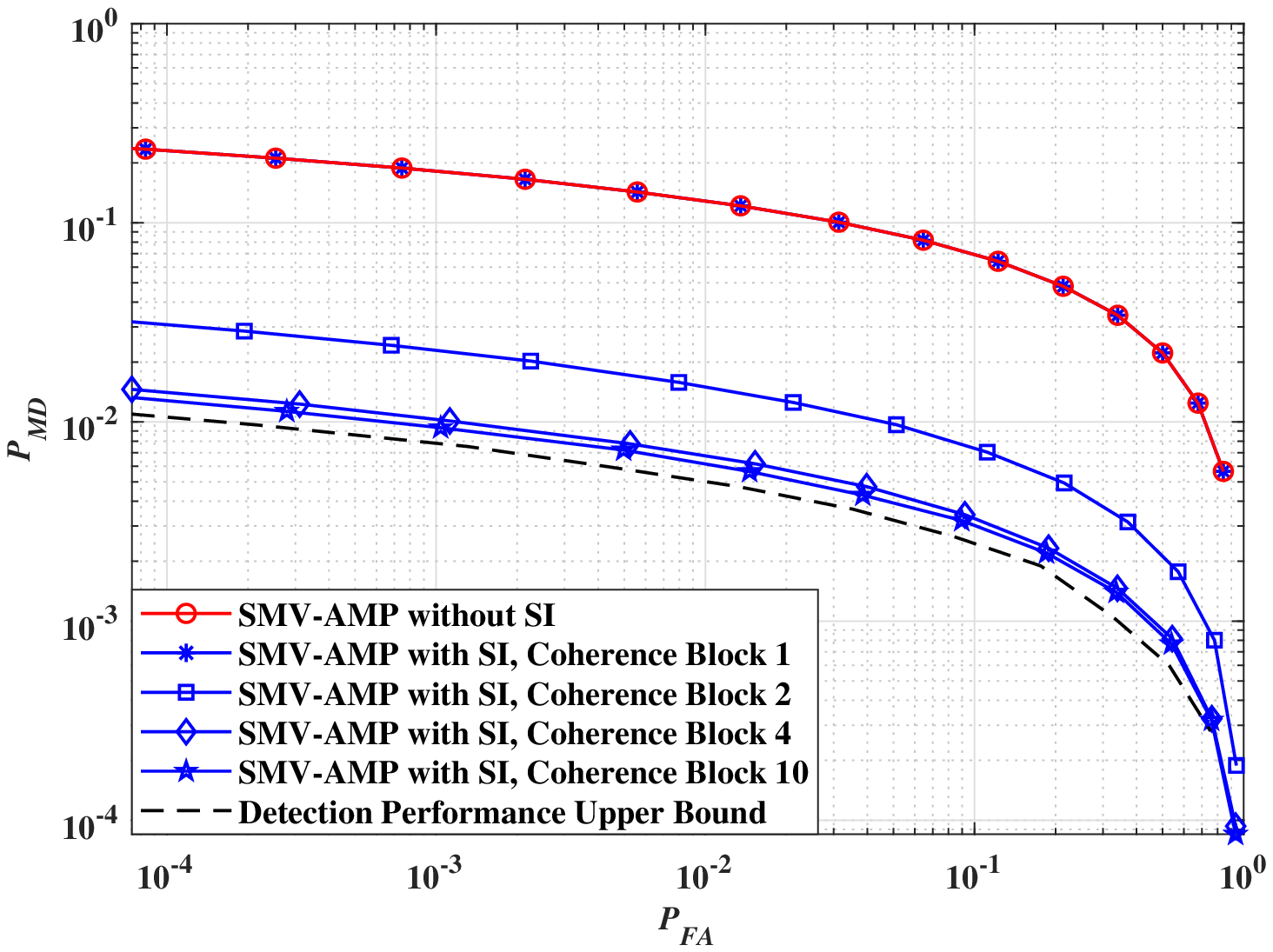}\label{AD_ST_M1}}
	\subfigure[Channel Estimation Performance]{\includegraphics[height=6cm]{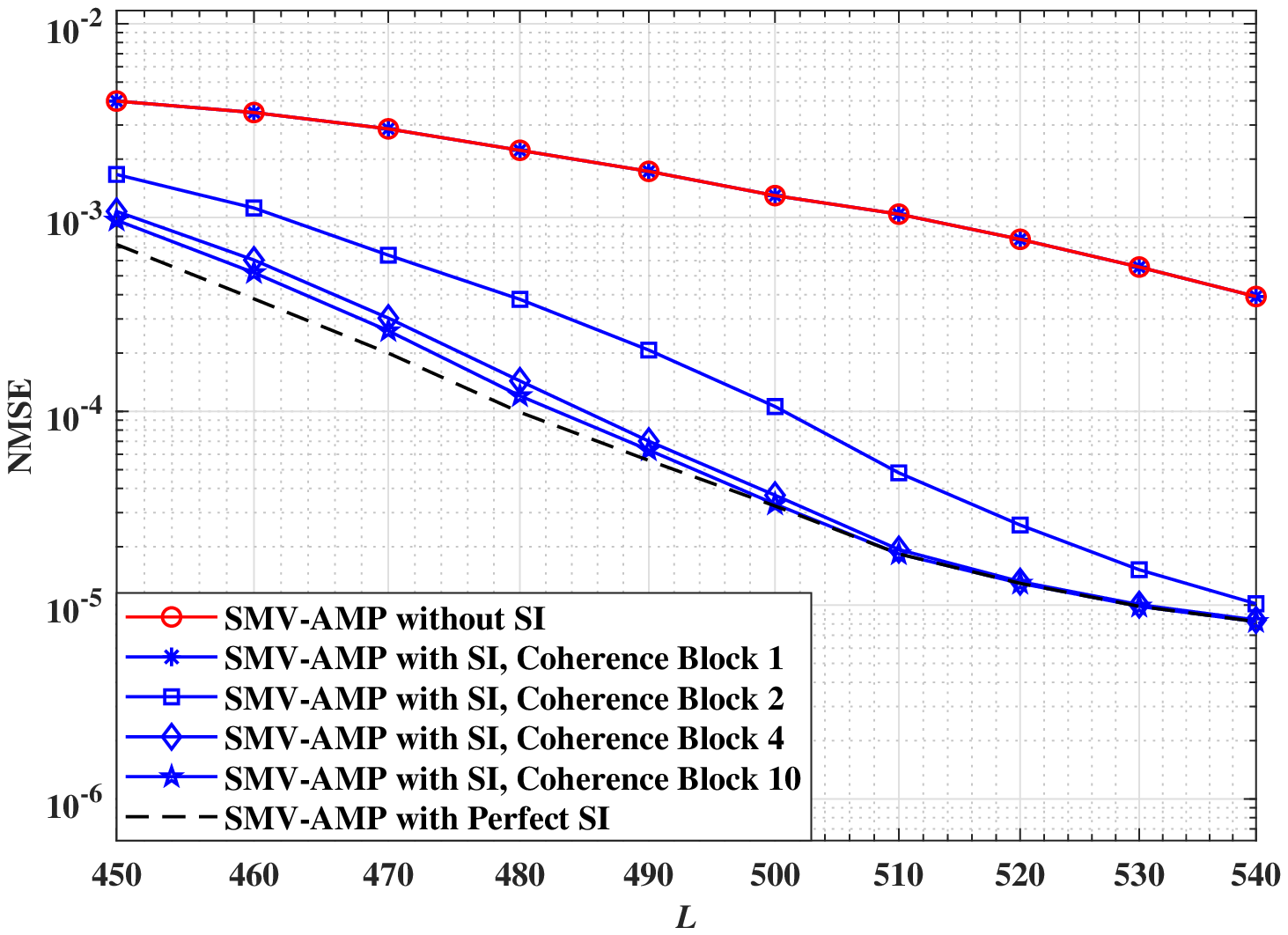}\label{CE_ST_M1}}
	\vspace{-3mm}
    \caption{Activity detection and channel estimation performance under SI-aided SMV-AMP framework with unknown channel distribution} \label{Fig_unknown_m1}
    \vspace{-6mm}
\end{figure}

To start with, we consider the case when the BS is equipped with one antenna, i.e., $M=1$. In this case, we term our proposed algorithm as SMV-AMP with SI. In Fig. \ref{AD_ST_M1}, we show the tradeoff between the probabilities of false alarm $P_{FA}$ and missed detection $P_{MD}$ under $L=500$, which is obtained by varying the value of the threshold $l$ in (\ref{llr_detection}). It is observed from Fig. \ref{AD_ST_M1} that our proposed framework outperforms the conventional AMP algorithm without utilizing SI. Moreover, under our proposed SI-aided AMP framework, the performance achieved in coherence block $10$ is much better than that achieved in coherence block $2$, which is better than that achieved in coherence block $1$. This shows that our proposed framework is capable of intelligently exploiting the SI obtained in the previous coherence block to improve the detection performance. At last, it is observed that after several coherence blocks, the performance of our proposed SI-aided AMP framework will converge very closely to the performance upper bound where the SI is perfect and the binary threshold is obtained using exhaustive search. This implies that although we design the binary threshold (\ref{wglasso}) for the approximate least-favorable channel distribution, it also works well under the Rayleigh fading channel.

Besides the device activity detection, we show in Fig. \ref{CE_ST_M1} the channel estimation performance under the Rayleigh fading channel model. For the performance metric, we define the normalized MSE for channel estimation in coherence block $j$ as
\begin{align}\label{normalizedmse}
    \text{NMSE}^{(j)}=\frac{{\sum_{n=1}^{N}\mathop{\mathbb{E}}[\|\hat{\boldsymbol{x}}_n^{(j)}-\boldsymbol{x}_n^{(j)}\|^2]}}{\sum_{n=1}^{N}\mathop{\mathbb{E}}[\|\boldsymbol{x}_n^{(j)}\|^2]}, \quad\forall j.
\end{align}
\begin{figure}[t]
	\centering
    \subfigure[Activity Detection Performance]{
    \label{AD_ST_M2}
    \includegraphics[width=0.47\columnwidth]{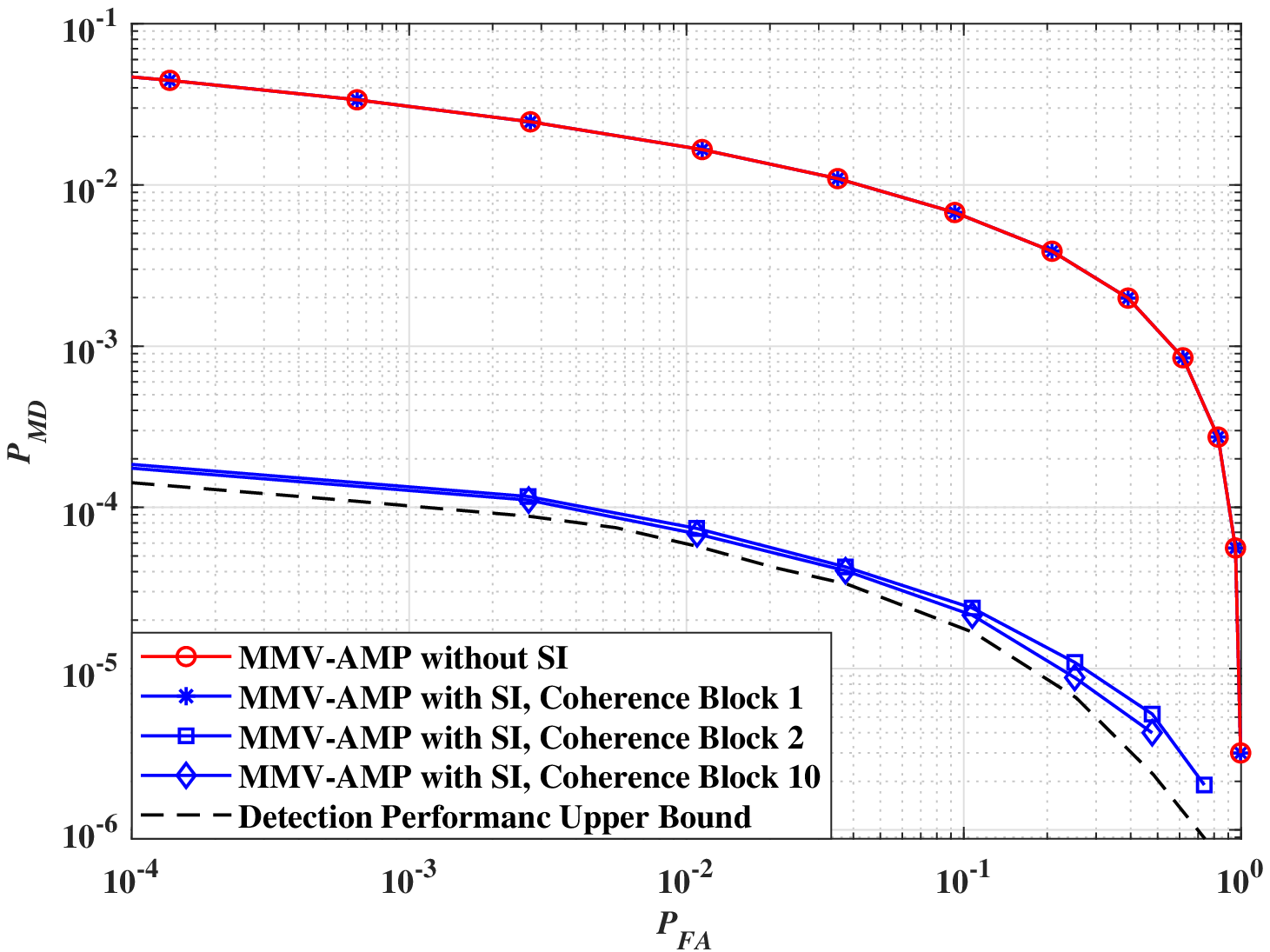}
    }
	\subfigure[Channel Estimation Performance]{
	\label{CE_ST_M2}
	\includegraphics[width=0.47\columnwidth]{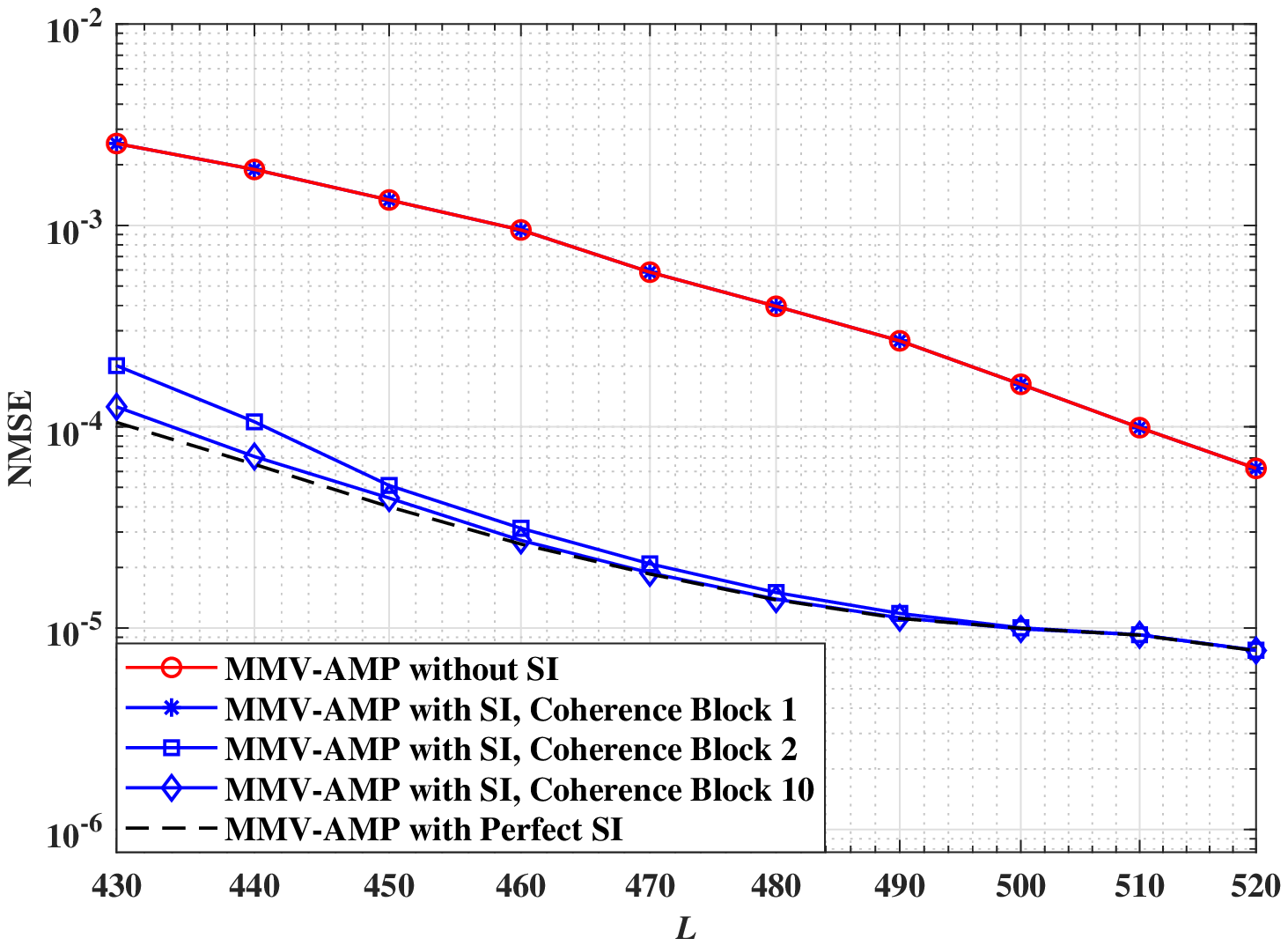}
	}
	\vspace{-3mm}
    \caption{Activity detection and channel estimation performance under SI-aided MMV-AMP framework with unknown channel distribution}\vspace{-10mm} \label{Fig_unknown_m2}
\end{figure}We consider a pilot sequence length ranging from $450$ to $540$. It is observed that the performance of our proposed SI-aided AMP framework will converge very closely to that achieved by the AMP algorithm with perfect SI, especially when the pilot sequence is long such that the imperfect SI can provide accurate information on the device activity in the previous coherence block.
\begin{figure}[t]
	\centering
	\subfigure[Activity Detection Performance]{
	\includegraphics[height=6cm]{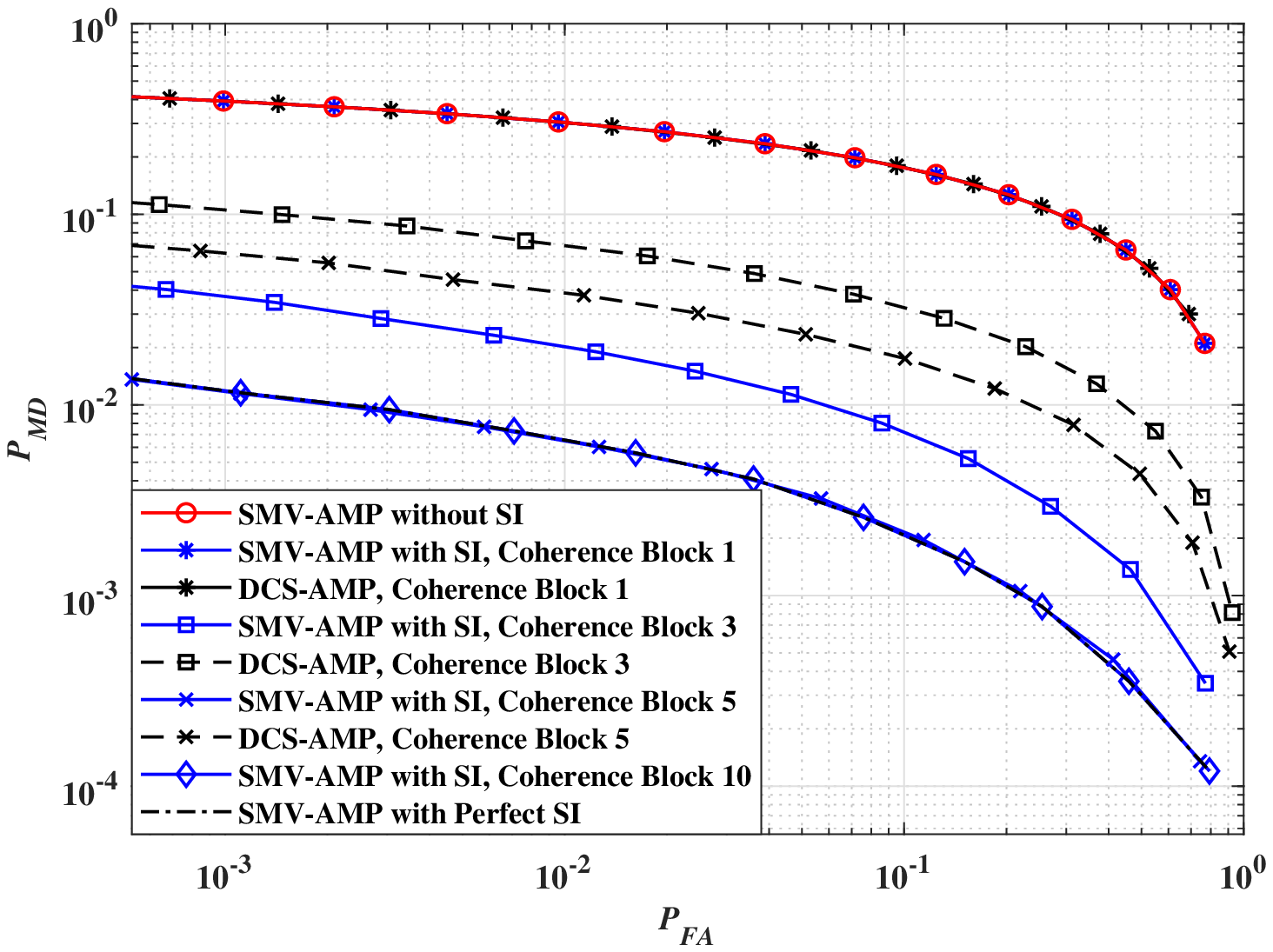}\label{AD_MMSE_M1}}
	\subfigure[Channel Estimation Performance]{\includegraphics[height=6cm]{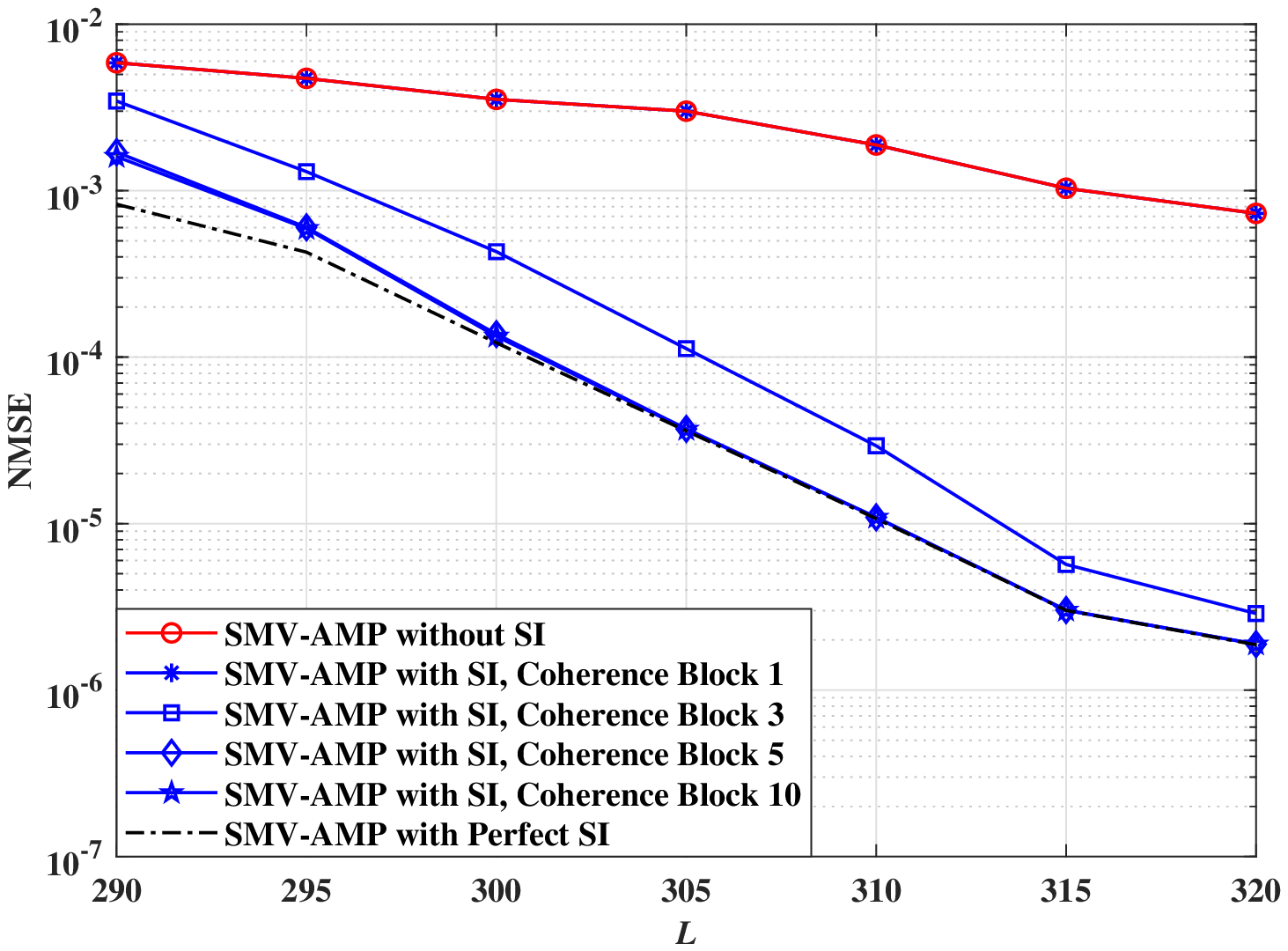}
	\label{CE_MMSE_M1}}\vspace{-3mm}
    \caption{Activity detection and channel estimation performance under SI-aided SMV-AMP framework with known channel distribution} \label{Fig_known_m1}
	\vspace{-6mm}
\end{figure}
\begin{figure}[t]
	\centering
	\subfigure[Activity Detection Performance]{
	\includegraphics[height=6cm]{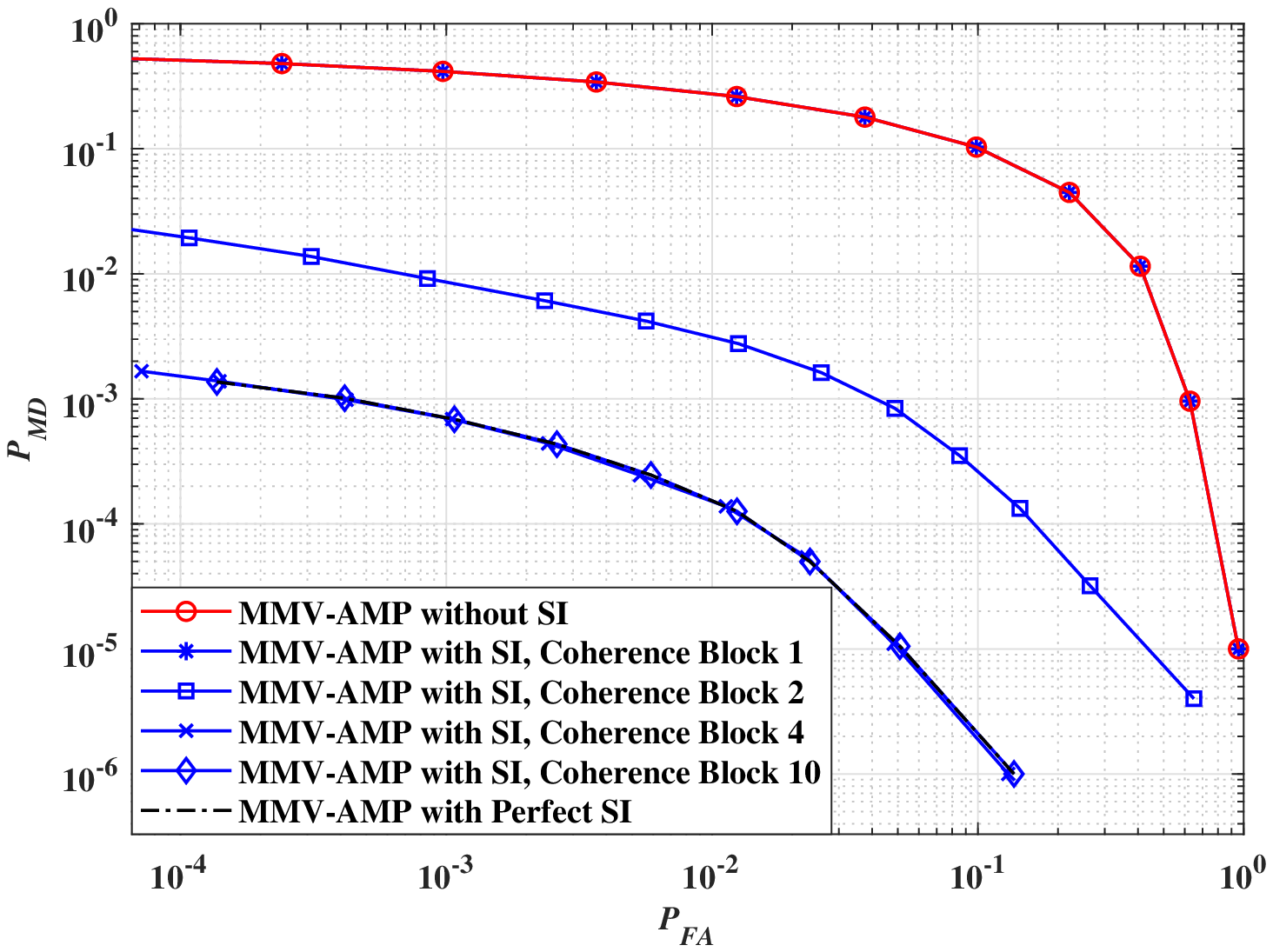}\label{AD_MMSE_M2}}
	\subfigure[Channel Estimation Performance]{\includegraphics[height=6cm]{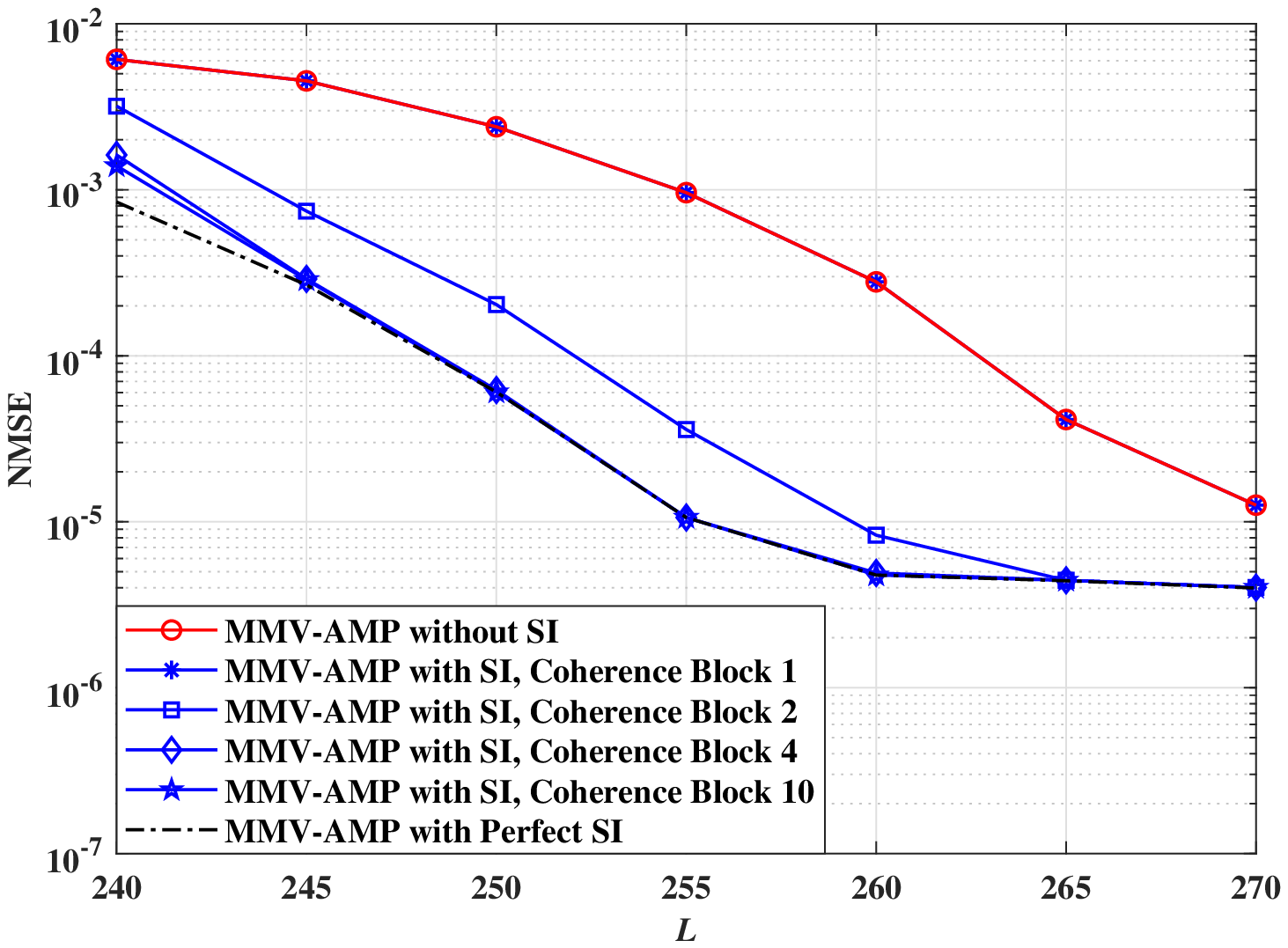}
	\label{CE_MMSE_M2}}	\vspace{-8mm}
    \caption{Activity detection and channel estimation performance under SI-aided MMV-AMP framework with known channel distribution}\label{Fig_known_m2}
	\vspace{-10mm}
\end{figure}

Secondly, we consider the case when the BS is equipped with multiple antennas with $M=2$. In this case, we term our proposed framework as MMV-AMP with SI. Figs. \ref{AD_ST_M2} and \ref{CE_ST_M2} show the tradeoff between the probabilities of false alarm $P_{FA}$ and missed detection $P_{MD}$ under $L=450$ and the normalized MSE with $L$ ranging from $430$ to $520$, respectively. It is observed that compared to the single-antenna case in Fig. \ref{Fig_unknown_m1}, it takes much fewer coherence blocks before the performance of the SI-aided MMV-AMP framework converges to a stable state. This is because with multiple antennas, the SI in the previous coherence block is more accurate, which can be utilized more efficiently in the current coherence block. Moreover, similar to the SMV case, our proposed SI-aided MMV-AMP framework significantly outperforms the MMV-AMP algorithm without using SI, and also performs closely to the MMV-AMP algorithm with perfect SI. The above results validate the effectiveness of our proposed framework with unknown channel distribution.

\subsection{Activity Detection and Channel Estimation Performance with Known Channel Distribution}

Next, we evaluate the performance of our proposed SI-aided MMV-AMP framework when the channel distribution is known. We consider three benchmark schemes. Besides the AMP algorithms with perfect SI or without SI, we also consider the dynamic compressed sensing via approximate message passing (DCS-AMP) algorithm proposed in \cite{DCS-AMP} as the benchmark scheme. The DCS-AMP is implemented in filtering model to match our setting. In Fig. \ref{AD_MMSE_M1} , we show the tradeoff between the probabilities of false alarm $P_{FA}$ and missed detection $P_{MD}$ when the BS is equipped with one antenna (i.e., $M=1$) and the length of the pilot sequence is set as $L=300$, while in Fig. \ref{CE_MMSE_M1}, we show the normalized MSE performance when $M=1$ and $L$ ranges from $290$ to $320$. It is observed that similar to the case without channel distribution information, when the Rayleigh fading channel model is known, the performance of our proposed SI-aided AMP algorithm is much better than that of the AMP algorithm without using SI, and very close to the AMP algorithm with perfect SI. It is also observed that the proposed SI-aided SMV-AMP algorithm outperforms the DCS-AMP algorithm significantly in coherence blocks $3$ and $5$, because our proposed scheme is built on the true statistical correlation between the estimation in the last coherence block and the user effective channels in the current coherence block. Similar observations are also observed from Fig. \ref{Fig_known_m2} when the BS is equipped with $M=2$ antennas, where the pilot sequence length is set as $L=250$ in Fig. \ref{AD_MMSE_M2} and that varies from $240$ to $270$ in Fig. \ref{CE_MMSE_M2}. Last, by comparing Fig. \ref{Fig_unknown_m1} and Fig. \ref{Fig_known_m1}, as well as Fig. \ref{Fig_unknown_m2} and Fig. \ref{Fig_known_m2}, it is observed that for both the single-antenna BS case and the multi-antenna BS case, if the channel distribution is known and utilized, the device activity detection and channel estimation performance can be significantly improved.

\section{Conclusions}\label{sec_conclusion}
In this paper, we proposed a comprehensive framework to utilize the temporal device activity correlation among adjacent blocks for device activity detection and channel estimation in massive IoT connectivity systems, in order to improve the detection and estimation performance with a small number of BS antennas. Specifically, we established a novel SI-aided MMV-AMP framework, where the estimation result in the previous coherence block is leveraged as SI for devising higher-quality denoisers in the current coherence block. For the case with unknown channel distribution information, we proposed a design for the SI-based soft-thresholding denoisers based on the least-favorable channel distribution. For the case with known channel distribution information, we derived the SI-based MMSE denoisers. Numerical results show that the proposed SI-aided MMV-AMP framework significantly improves the detection and estimation performance compared to the conventional MMV-AMP algorithms without exploiting SI, which validates the effectiveness of our proposed framework in smartly exploiting SI. Therefore, the proposed framework provides a promising approach to achieve high-quality activity detection and channel estimation even with a small number of antennas at the BS.

\begin{appendices}
\section{Proof of Theorem \ref{theorem1}}\label{worstcasemse}
In this proof, for simplicity, we omit the subscript $n$ in the notations. First, the probability of the event $\delta^{(j)}=0$ and $\hat{\delta}^{(j-1)}=1$ is
\begin{align}\label{01}
    Pr(\delta^{(j)}=0,\hat{\delta}^{(j-1)}=1)&={Pr(\delta^{(j)}=0,\hat{\delta}^{(j-1)}=1,\delta^{(j-1)}=0)+Pr(\delta^{(j)}=0,\hat{\delta}^{(j-1)}=1,\delta^{(j-1)}=1)}\notag\\
    &\overset{(a)}{=}{Pr(\delta^{(j)}=0\mid \delta^{(j-1)}\!=\!0)Pr(\hat{\delta}^{(j-1)}=1\mid \delta^{(j-1)}=0)Pr(\delta^{(j-1)}=0)}\notag\\
    &~~~+{Pr(\delta^{(j)}=0\mid \delta^{(j-1)}=1)Pr(\hat{\delta}^{(j-1)}=1\mid \delta^{(j-1)}=1)Pr(\delta^{(j-1)}=1)}\notag\\
    &\overset{(b)}{=}(1-\beta)(1-\lambda)\frac{\overline{\Gamma}\big(M, (l^{\left(j-1\right)})^{2} (\tau_{\infty}^{\left(j-1\right)})^{-2}\big)}{\Gamma(M)}+{(1-\alpha)\lambda},\quad \forall j,
\end{align}where $(a)$ is due to the fact that $\delta^{(j)}$ and $\hat{\delta}^{(j-1)}$ are independent given $\delta^{(j-1)}$ \cite{conditional_independence}, and $(b)$ holds because given the approximate least-favorable distribution (\ref{each_least_favorable}), we have
\begin{align}\label{P_fa_proof}
    &Pr(\hat{\delta}^{(j-1)}=1\mid \delta^{(j-1)}=0)=Pr(\|\boldsymbol{V}^{(j)} \| > \frac{l^{(j-1)}}{\tau_{\infty}^{(j-1)}}){=}\frac{\overline{\Gamma}\big(M, (l^{\left(j-1\right)})^{2} (\tau_{\infty}^{\left(j-1\right)})^{-2}\big)}{\Gamma(M)},\quad\forall j,\\
    &Pr(\hat{\delta}^{(j-1)}=1\mid \delta^{(j-1)}=1)=1,\quad\forall j.
\end{align}Moreover, when $\delta^{(j)}=0$ and $\hat{\delta}^{(j-1)}=1$, the channel estimation MSE is
\begin{align}\label{expcaseii}
    &\mathop{\mathbb{E}}\big [ \| \text{g}_t^{(j)}(\tau_t^{\left(j\right)}\boldsymbol{V}^{(j)},\theta_{1,t}^{\left(j\right)}) \|^2 \big]=\int_{\|\boldsymbol{V}^{\left(j\right)} \|>\frac{\theta_{1,t}^{\left(j\right)}}{\tau_t^{(j)}}}\big(\|\tau_t^{(j)}\boldsymbol{V}^{\left(j\right )} \|-{\theta_{1,t}^{\left(j\right)}}\big)^2p_{\|\boldsymbol{V}^{\left(j\right)}\|} {\rm{d}}\|\boldsymbol{V}^{(j)}\|\notag\\
    =&\frac{({\theta_{1,t}^{\left(j\right)}})^2\overline{\Gamma}\big(M,(\frac{\theta_{1,t}^{\left(j\right)}}{\tau_t^{(j)}})^2\big)-2{\theta_{1,t}^{\left(j\right)}}{\tau_t^{(j)}}\overline{\Gamma}\big(M+\frac{1}{2},(\frac{\theta_{1,t}^{\left(j\right)}}{\tau_t^{(j)}})^2\big)+(\tau_t^{(j)})^2\overline{\Gamma}\big(M+1,(\frac{\theta_{1,t}^{\left(j\right)}}{\tau_t^{(j)}})^2\big)}{\Gamma(M)},\quad \forall j,t.
\end{align}

Next, consider the event that $\delta^{(j)}=1$ and $\hat{\delta}^{(j-1)}=1$. It can be shown that the probability of the above event is
\begin{align}\label{11}
    Pr(\delta^{(j)}=1,\hat{\delta}^{(j-1)}=1)&=\beta(1-\lambda)\frac{\overline{\Gamma}\big(M, (l^{\left(j-1\right)})^{2} (\tau_{\infty}^{\left(j-1\right)})^{-2}\big)}{\Gamma(M)}+\alpha\lambda,\quad \forall j.
\end{align}Moreover, the channel estimation MSE is
\begin{align}
    &\mathop{\mathbb{E}}[\|g_t^{(j)}(\boldsymbol{h}^{\left(j\right)}+ \tau^{\left(j\right)}\boldsymbol{V}^{(j)},\theta_{1,t}^{(j)}) -\boldsymbol{h}^{\left(j\right)} \|^2]=\mathop{\mathbb{E}}_{\boldsymbol{h}^{(j)}\sim\boldsymbol{\hat{\mu}}_{n,M}^{\ast}}R_t^{(j)}(\|\boldsymbol{h}^{(j)}\|,\theta_{1,t}^{(j)})\notag\\&=\sum_{m=1}^{M}\mathop{\mathbb{E}}_{\boldsymbol{v}_m^{(j)}}\big[\|\tau_t^{(j)}\boldsymbol{v}_m^{(j)}-\frac{\theta_{1,t}^{(j)}}{\sqrt{M}}\|^2\big]=(\tau_t^{(j)})^2M+(\theta_{1,t}^{(j)})^2, \quad\forall j,t.\label{exp_casei}
\end{align}

Third, consider the event that $\delta^{(j)}=0$ and $\hat{\delta}^{(j-1)}=0$. It can be shown that the probability of the above event is
\begin{align}\label{00}
    Pr(\delta^{(j)}=0,\hat{\delta}^{(j-1)}=0)=(1-\beta)(1-\lambda)(1-\frac{\overline{\Gamma}\big(M, (l^{\left(j-1\right)})^{2} (\tau_{\infty}^{\left(j-1\right)})^{-2}\big)}{\Gamma(M)}),\quad \forall j.
\end{align}Moreover, the channel estimation MSE
$\mathop{\mathbb{E}}\big [ \| g_t^{(j)}(\tau_t^{\left(j\right)}\boldsymbol{V}^{(j)},\theta_{2,t}^{\left(j\right)}) \|^2 \big]$ has the same form as (\ref{expcaseii}) but with $\theta_{1,t}^{\left(j\right)}$ replaced by $\theta_{2,t}^{\left(j\right)}$, $\forall j,t$.

Last, consider the event that $\delta^{(j)}=1$ and $\hat{\delta}^{(j-1)}=0$. It can be shown that the probability of the above event and the channel estimation MSE are respectively expressed as
\begin{align}\label{10}
    &Pr(\delta^{(j)}=1,\hat{\delta}^{(j-1)}=0)=\beta(1-\lambda)(1-\frac{\overline{\Gamma}\big(M, (l^{\left(j-1\right)})^{2} (\tau_{\infty}^{\left(j-1\right)})^{-2}\big)}{\Gamma(M)}), \quad\forall j,\\
    &\mathop{\mathbb{E}}\big[\| g_t^{(j)}(\boldsymbol{h}^{\left(j\right)}+ \tau_t^{\left(j\right)}\boldsymbol{V}^{(j)},\theta_{2, t}^{(j)}) -\boldsymbol{h}^{\left(j\right)} \|^2 \big]=(\tau_t^{(j)})^2M\!+\!({\theta_{2,t}^{\left(j\right)}})^2,\quad \forall j,t.\label{exp_casev}
\end{align}
Thus, by plugging (\ref{01}), (\ref{expcaseii})-(\ref{exp_casev}) into (\ref{iv_mse_2}), Theorem \ref{theorem1} is proved.

\section{Proof of Theorem \ref{theorem2}}\label{threshold}
In this proof, we omit the subscripts $t$ and $n$ in all the notations for simplicity. The first derivative of the upper incomplete gamma function $\overline{\Gamma}(M,x)$ with respect to $x$ is given as $\frac{\partial{\overline{\Gamma}(M,x)}}{\partial x}=-x^{M-1}e^{-x}$. Then, the first derivatives of $\mathrm{MSE}_{i}^{(j)}(\theta_{i}^{(j)}, \boldsymbol{\hat{\mu}}^{\ast})$ with respect to $\theta_i^{(j)}$, $i=1,2$, are given as
\begin{align}\label{first_deriv_1}
    \frac{\partial{{\rm {MSE}}}_i^{(j)}(\theta_i^{(j)},\boldsymbol{\hat{\mu}}^{\ast})}{\partial  (\theta_i^{(j)})}=2f_i(\theta_i^{(j)}),\quad i=1,2, \forall j,
\end{align}where $f_1(\theta_1^{(j)})$ and $f_2(\theta_2^{(j)})$ are given in (\ref{free_par1}) and (\ref{free_par2}), respectively. Moreover, the second derivatives of ${\text{MSE}}_i^{(j)}(\theta_{i}^{(j)},\boldsymbol{\hat{\mu}}^{\ast})$ with respect to $\theta_i^{(j)}$, $i=1,2$, are given as
\begin{align}
    \frac{\partial^2\mathrm{MSE}_1^{(j)}(\theta_{1}^{(j)}, \boldsymbol{\hat{\mu}}^{\ast})}{\partial (\theta_1^{(j)})^2}=2f_1^{'}(\theta_1^{(j)})&=2[(1-\beta)(1-\lambda)\varsigma_{\infty}^{(j-1)}+(1-\alpha)\lambda]\frac{\overline{\Gamma}\big(M,(\frac{\theta_1^{\left(j\right)}}{\tau^{(j)}})^2\big)}{\Gamma\left(M\right)}\notag\\&+2[\beta(1-\lambda)\varsigma_{\infty}^{(j-1)}+\alpha\lambda]>0,\quad\forall j,\\
    \frac{\partial^2\mathrm{MSE}_2^{(j)}( \theta_{2}^{(j)},\boldsymbol{\hat{\mu}}^{\ast})}{\partial (\theta_2^{(j)})^2}=2f_2^{'}(\theta_2^{(j)})&=2(1-\beta)(1-\lambda)(1-\varsigma_{\infty}^{(j-1)})\frac{\overline{\Gamma}\big(M,(\frac{\theta_2^{\left(j\right)}}{\tau^{(j)}})^2\big)}{\Gamma\left(M\right)}\notag\\&+2\beta(1-\lambda)(1-\varsigma_{\infty}^{(j-1)})>0, \quad\forall j,
\end{align}where $\varsigma_{\infty}^{(j-1)}$ is given in (\ref{Th1_P_fa}). Thus, $\mathrm{MSE}_i^{(j)}(\theta_{i}^{(j)},\boldsymbol{\hat{\mu}}^{\ast})$ is convex with respect to $\theta_i^{\left(j\right)}$, $i=1,2$, $\forall j$. To minimize the convex functions $\mathrm{MSE}_i^{(j)}(\theta_{i}^{(j)},\boldsymbol{\hat{\mu}}^{\ast})$, $i=1,2$, $\forall j$, in the following, we show that there always exist a positive solution $\hat{\theta}_1^{(j),\ast}$ to the equation $f_1(\theta_1^{(j)})=0$ and a positive solution $\hat{\theta}_2^{(j),\ast}$ to the equation $f_2(\theta_2^{(j)})=0$. First, it can be shown that
\begin{align}\label{f1zero}
f_1(0)=-[(1-\beta)(1-\lambda)\varsigma_{\infty}^{(j-1)}+(1-\alpha)\lambda]\frac{\tau^{(j)}\Gamma(M+\frac{1}{2})}{\Gamma(M)}<0,\quad \forall j.
\end{align} Next, it can be shown that
\begin{align}\label{f1infty}
    f_1(\infty)\!&=\![(1\!-\!\beta)(1\!-\!\lambda)\varsigma_{\infty}^{(j-1)}\!+\!(1\!-\!\alpha)\lambda]\!\lim_{{\theta}_{1}^{(j)} \to \infty} \xi^{\left(j\right)}({\theta}_{1}^{(j)})\!+\![\beta(1\!-\!\lambda)\varsigma_{\infty}^{(j-1)}+\alpha\lambda]\!\lim_{{\theta}_{1}^{(j)} \to \infty}{\theta}_{1}^{(j)}\notag\\
    &=\![(1\!-\!\beta)(1\!-\!\lambda)\varsigma_{\infty}^{(j-1)}\!+\!(1\!-\!\alpha)\lambda]\!\lim_{{\theta}_{1}^{(j)} \to \infty}\frac{{\theta}_{1}^{(j)}\overline{\Gamma}\big(M,(\frac{{\theta}_{1}^{(j)}}{\tau_t^{(j)}})^2\big)}{\Gamma(M)} \!+\![\beta(1\!-\!\lambda)\varsigma_{\infty}^{(j-1)}+\alpha\lambda]\!\lim_{{\theta}_{1}^{(j)} \to \infty}{\theta}_{1}^{(j)}\notag\\
    &\overset{(c)}{>}0,\quad \forall j,
\end{align}where $(c)$ holds because $\lim_{{\theta}_{1}^{(j)} \to \infty}\frac{{\theta}_{1}^{(j)}\overline{\Gamma}\big(M,(\frac{{\theta}_{1}^{(j)}}{\tau_t^{(j)}})^2\big)}{\Gamma(M)}$ is no smaller than $0$. Since $f_1(\theta_1^{(j)})$ is an increasing function over $\theta_1^{(j)}$, (\ref{f1zero}) and (\ref{f1infty}) indicate that there is a positive solution $\hat{\theta}_1^{(j),\ast}$ such that $f_1(\hat{\theta}_1^{(j),\ast})=0$. Similarly, it can be shown that there is a positive solution $\hat{\theta}_2^{(j),\ast}$ such that $f_2(\hat{\theta}_2^{(j),\ast})=0$. Theorem \ref{theorem2} is thus proved.

\section{Proof of Lemma \ref{lemma_appc}}\label{app_mmse_derivation}

In this proof, for simplicity, we omit the subscripts $t$ and $n$ in the all the notations. In the MMSE denoiser (\ref{expformmse}), the conditional expectation can be given by
\begin{align}\label{expectation}
    \mathop{\mathbb{E}}[\boldsymbol{X}^{\left(j\right)}\mid \boldsymbol{\tilde x}^{\left(j\right)}, \boldsymbol{\tilde x}_\infty^{\left(j-1\right)}]&
    \overset{{(d)}}{=}\mathop{\mathbb{E}}[\boldsymbol{X}^{\left(j\right)} \mid \boldsymbol{\tilde x}^{\left(j\right)}, \boldsymbol{\tilde{x}}_\infty^{(j-1)},Case ~~1]p( {Case ~~1}\mid \boldsymbol{\tilde x}^{\left(j\right)}, \boldsymbol{\tilde x}_\infty^{\left(j-1\right)})\notag\\
    &+\mathop{\mathbb{E}}[\boldsymbol{X}^{\left(j\right)} \mid \boldsymbol{\tilde x}^{\left(j\right)}, \boldsymbol{\tilde x}_\infty^{\left(j-1\right)},Case~~ 3]p( {Case ~~3}\mid \boldsymbol{\tilde x}^{\left(j\right)}, \boldsymbol{\tilde{x}}_\infty^{\left(j-1\right)}),\quad \forall j,
\end{align}
where $(d)$ holds because $\boldsymbol{X}^{\left(j\right)}=\boldsymbol{0}$ for \emph{Case 2} and \emph{Case 4} according to Section \ref{sec_sys}. In the following, we focus on \emph{Case 1} and \emph{Case 3} to characterize (\ref{expectation}).

\emph{Case 1}:
According to Section \ref{sec_sys}, under \emph{Case 1}, it follows that $\boldsymbol{x}^{\left(j-1\right)}=\boldsymbol{h}^{\left(j-1\right)}$ and $\boldsymbol{x}^{\left(j\right)}=\boldsymbol{h}^{\left(j\right)}$. Based on (\ref{model}) and (\ref{state_estimation}), we have $\boldsymbol{\tilde x}^{\left(j\right)}=\boldsymbol{h}^{\left(j\right)}+(\boldsymbol{\Sigma}^{\left(j\right)})^\frac{1}{2}\boldsymbol{V}, \boldsymbol{\tilde x}_{\infty}^{\left(j-1\right)}=\boldsymbol{h}^{\left(j-1\right)}+(\boldsymbol{\Sigma} _\infty^{\left(j-1\right)})^\frac{1}{2}\boldsymbol{V}$. Moreover, due to the independence among variables $\boldsymbol{h}^{\left(j\right)}$, $\boldsymbol{h}^{\left(j-1\right)}$, and $(\boldsymbol{\Sigma} _\infty^{\left(j-1\right)})^\frac{1}{2}\boldsymbol{V}$, $\mathop{\mathbb{E}}[\boldsymbol{X}^{\left(j\right)} \mid \boldsymbol{\tilde x}^{\left(j\right)}, \boldsymbol{\tilde{x}}_{\infty}^{\left(j-1\right)},Case ~~ 1]$ in (\ref{expectation}) can be given by
\begin{align}\label{exp_case1}
    \mathop{\mathbb{E}}[\boldsymbol{X}^{\left(j\right)} \mid \boldsymbol{\tilde x}^{\left(j\right)}, \boldsymbol{\tilde{x}}_{\infty}^{\left(j-1\right)},Case ~~ 1]=\mathop{\mathbb{E}}[\boldsymbol{h}^{\left(j\right)} \mid \boldsymbol{\tilde x}^{\left(j\right)}=\boldsymbol{h}^{\left(j\right)}+(\boldsymbol{\Sigma }^{\left(j\right)})^\frac{1}{2}\boldsymbol{V}],\quad \forall j.
\end{align} Next, $p({Case ~ 1}\mid \boldsymbol{\tilde x}^{\left(j\right)}, \boldsymbol{\tilde{x}}_\infty^{\left(j-1\right)})$ in (\ref{expectation}) is calculated as
\begin{align}\label{pcase1}
    p( {Case ~~1}\mid \boldsymbol{\tilde x}^{\left(j\right)},\boldsymbol{\tilde x}_{\infty}^{\left(j-1\right)})&=\frac{P(Case1)p(\boldsymbol{\tilde x}^{\left(j\right)}, \boldsymbol{\tilde{x}}_{\infty}^{\left(j-1\right)}\mid Case~~1)}{p(\boldsymbol{\tilde x}^{\left(j\right)}, \boldsymbol{\tilde{x}}_{\infty}^{\left(j-1\right)})}\notag \\
    &=\frac{\alpha\lambda\boldsymbol{\psi}_{\gamma\boldsymbol{I}+\boldsymbol{\Sigma}^{\left(j\right)}}(\boldsymbol{\tilde x}^{\left(j\right)})\boldsymbol{\psi}_{\gamma\boldsymbol{I}+\boldsymbol{\Sigma} _\infty^{\left(j-1\right)}}(\boldsymbol{\tilde{x}}_{\infty}^{\left(j-1\right)})}{p(\boldsymbol{\tilde x}^{\left(j\right)}, \boldsymbol{\tilde{x}}_{\infty}^{\left(j-1\right)})}, \quad\forall j,
\end{align}
where $\boldsymbol{\psi}_{\sigma^2\boldsymbol{I}}(\boldsymbol{x})=\frac{1}{\pi\left |(\sigma^2\boldsymbol{I}) \right | }exp({-\boldsymbol{x}^H{(\sigma ^2\boldsymbol{I})^{-1}\boldsymbol{x})}}$ is the PDF of multivariate complex Gaussian distribution. We will derive the joint probability, i.e., $p(\boldsymbol{\tilde x}^{\left(j\right)},\boldsymbol{\tilde{x}}_{\infty}^{\left(j-1\right)})$ later.

\emph{Case 3}:
Similar to \emph{Case 1}, the conditional expectation can be calculated by:
\begin{equation}\label{ecase3}
    \mathop{\mathbb{E}}[\boldsymbol{X}^{\left(j\right)} \mid \boldsymbol{\tilde x}^{\left(j\right)}, \boldsymbol{\tilde{x}}_{\infty}^{\left(j-1\right)},Case~~3]=\mathop{\mathbb{E}}[\boldsymbol{h}^{\left(j\right)} \mid \boldsymbol{\tilde x}^{\left(j\right)}=\boldsymbol{h}^{\left(j\right)}+(\boldsymbol{\Sigma }^{\left(j\right)})^\frac{1}{2}\boldsymbol{V}],\quad \forall j.
\end{equation}
Moreover, similar to (\ref{pcase1}),
\begin{align}\label{pcase3}
    p( {Case 3}\mid \boldsymbol{\tilde x}^{\left(j\right)}, \boldsymbol{\tilde{x}}_{\infty}^{\left(j-1\right)})=\frac{\beta(1-\lambda)\boldsymbol{\psi}_{\gamma\boldsymbol{I}+\boldsymbol{\Sigma}^{\left(j\right)}}(\boldsymbol{\tilde x}^{\left(j\right)})\boldsymbol{\psi}_{\boldsymbol{\Sigma} _\infty^{\left(j-1\right)}}(\boldsymbol{\tilde{x}}_{\infty}^{\left(j-1\right)}) }{p(\boldsymbol{\tilde x}^{\left(j\right)},\boldsymbol{\tilde{x}}_{\infty}^{\left(j-1\right)})},\quad \forall j.
\end{align}

To derive $p({Case ~~ 1}\mid \boldsymbol{\tilde x}^{\left(j\right)},\boldsymbol{\tilde x}_\infty^{\left(j-1\right)})$ in (\ref{pcase1}) and $p( {Case ~~ 3}\mid \boldsymbol{\tilde x}^{\left(j\right)}, \boldsymbol{\tilde{x}}_{\infty}^{\left(j-1\right)})$ in (\ref{pcase3}), the last step is to characterize $p(\boldsymbol{\tilde x}^{\left(j\right)}, \boldsymbol{\tilde{x}}_\infty^{\left(j-1\right)})$. Similar to $p(\boldsymbol{\tilde x}^{\left(j\right)}, \boldsymbol{\tilde{x}}_\infty^{\left(j-1\right)}, Case ~ 1)$ shown in (\ref{pcase1}) and $p(\boldsymbol{\tilde x}^{\left(j\right)}, \boldsymbol{\tilde{x}}_\infty^{\left(j-1\right)}, Case ~ 3)$ shown in (\ref{pcase3}), it can be shown that
\begin{align}
    &p(\boldsymbol{\tilde x}^{\left(j\right)}, \boldsymbol{\tilde{x}}_{\infty}^{\left(j-1\right)},Case~~2)=(1-\alpha)\lambda \boldsymbol{\psi}_{\boldsymbol{\Sigma}^{\left(j\right)}}(\boldsymbol{\tilde x}^{\left(j\right)})\boldsymbol{\psi}_{\gamma\boldsymbol{I}+\boldsymbol{\Sigma} _\infty^{\left(j-1\right)}}(\boldsymbol{\tilde{x}}_{\infty}^{\left(j-1\right)}),\quad \forall j,\\
    &p(\boldsymbol{\tilde x}^{\left(j\right)}, \boldsymbol{\tilde{x}}_{\infty}^{\left(j-1\right)},Case~~4)=(1-\beta)(1-\lambda) \boldsymbol{\psi}_{\boldsymbol{\Sigma}^{\left(j\right)}}(\boldsymbol{\tilde x}^{\left(j\right)})\boldsymbol{\psi}_{\boldsymbol{\Sigma} _\infty^{\left(j-1\right)}}(\boldsymbol{\tilde{x}}_{\infty}^{\left(j-1\right)}),\quad \forall j.
\end{align}
Thus,
\begin{align}\label{totalp}
    &p(\boldsymbol{\tilde x}^{\left(j\right)}, \boldsymbol{\tilde{x}}_{\infty}^{\left(j-1\right)})= \sum_{i=1}^{4} p(\boldsymbol{\tilde x}^{\left(j\right)}, \boldsymbol{\tilde{x}}_{\infty}^{\left(j-1\right)}, Case~~i)\notag\\
    =&\alpha\lambda\boldsymbol{\psi}_{\gamma\boldsymbol{I}+\boldsymbol{\Sigma }^{\left(j\right)}}(\boldsymbol{\tilde x}^{\left(j\right)})\boldsymbol{\psi}_{\gamma\boldsymbol{I}+\boldsymbol{\Sigma} _\infty^{\left(j-1\right)}}(\boldsymbol{\tilde{x}}_{\infty}^{\left(j-1\right)})+(1-\alpha)\lambda \boldsymbol{\psi}_{\boldsymbol{\Sigma}^{\left(j\right)}}(\boldsymbol{\tilde x}^{\left(j\right)})\boldsymbol{\psi}_{\gamma\boldsymbol{I}+\boldsymbol{\Sigma}_\infty^{\left(j-1\right)}}(\boldsymbol{\tilde{x}}_{\infty}^{\left(j-1\right)})\notag\\
    +&\beta(1-\lambda)\boldsymbol{\psi}_{\gamma\boldsymbol{I}+\boldsymbol{\Sigma }^{\left(j\right)}}(\boldsymbol{\tilde x}^{\left(j\right)})\boldsymbol{\psi}_{\boldsymbol{\Sigma} _\infty^{\left(j-1\right)}}(\boldsymbol{\tilde{x}}_{\infty}^{\left(j-1\right)})+(1-\beta)(1-\lambda) \boldsymbol{\psi}_{\boldsymbol{\Sigma}^{\left(j\right)}}(\boldsymbol{\tilde x}^{\left(j\right)})\boldsymbol{\psi}_{\boldsymbol{\Sigma} _\infty^{\left(j-1\right)}}(\boldsymbol{\tilde{x}}_{\infty}^{\left(j-1\right)}),\quad \forall j,
\end{align}
Finally,
\begin{align}\label{mmse_cov}
    \mathop{\mathbb{E}}[\boldsymbol{X}^{\left(j\right)}\mid \boldsymbol{\tilde x}^{\left(j\right)}, \boldsymbol{\tilde x}_\infty^{\left(j-1\right)}] &= \frac{p(\boldsymbol{\tilde x}^{\left(j\right)}, \boldsymbol{\tilde{x}}_{\infty}^{\left(j-1\right)},Case 1)+p(\boldsymbol{\tilde x}^{\left(j\right)}, \boldsymbol{\tilde{x}}_{\infty}^{\left(j-1\right)},Case 3)}{p(\boldsymbol{\tilde{x}}^{\left(j\right)},\boldsymbol{\tilde{x}}_{\infty}^{\left(j-1\right)})}\mathop{\mathbb{E}}[\boldsymbol{h}^{\left(j\right)} \mid \boldsymbol{\tilde x}^{\left(j\right)}=\boldsymbol{h}^{\left(j\right)}+(\boldsymbol{\Sigma }^{\left(j\right)})^\frac{1}{2}\boldsymbol{V}]\notag\\
    &\overset{(e)}{=}\phi(\boldsymbol{\tilde x}^{\left(j\right)},\boldsymbol{\tilde x}_{\infty}^{\left(j-1\right)}){\gamma}({\gamma\boldsymbol{I}+\boldsymbol{\Sigma}^{\left(j\right)}})^{-1}{\boldsymbol{\tilde x}^{\left(j\right)}},\quad \forall j,
\end{align}
where
\begin{equation}\label{phi}
    \phi(\boldsymbol{\tilde x}^{\left(j\right)},\boldsymbol{\tilde x}_{\infty}^{\left(j-1\right)})=\frac{p(\boldsymbol{\tilde x}^{\left(j\right)}, \boldsymbol{\tilde{x}}_{\infty}^{\left(j-1\right)},Case 1)+p(\boldsymbol{\tilde x}^{\left(j\right)}, \boldsymbol{\tilde{x}}_{\infty}^{\left(j-1\right)},Case 3)}{p(\boldsymbol{\tilde{x}}^{\left(j\right)},\boldsymbol{\tilde{x}}_{\infty}^{\left(j-1\right)})},\quad \forall j.
\end{equation}
and $(e)$ is by standard estimation theory.

Similarly, using standard estimation theory, it can be shown that
\begin{align}
    &\mathop{\mathbb{E}}[\boldsymbol{X}^{\left(j\right)}{(\boldsymbol{X}^{\left(j\right)})}^{H}\mid \boldsymbol{\tilde{X}}^{\left(j\right)}=\boldsymbol{\tilde x}^{\left(j\right)}, \boldsymbol{\tilde{X}}_\infty^{\left(j-1\right)}=\boldsymbol{\tilde x}_\infty^{\left(j-1\right)}] \notag\\
    =&\phi(\boldsymbol{\tilde x}^{\left(j\right)},\boldsymbol{\tilde x}_{\infty}^{\left(j-1\right)})\mathop{\mathbb{E}}[\boldsymbol{h}^{\left(j\right)}(\boldsymbol{h}^{\left(j\right)})^H \mid \boldsymbol{\tilde x}^{\left(j\right)}=\boldsymbol{h}^{\left(j\right)}+(\boldsymbol{\Sigma }^{\left(j\right)})^\frac{1}{2}\boldsymbol{V}]
    \notag\\=& \phi(\boldsymbol{\tilde x}^{\left(j\right)},\boldsymbol{\tilde x}_{\infty}^{\left(j-1\right)})\big({\gamma}\boldsymbol{I}-\gamma^2(\gamma\boldsymbol{I}+\boldsymbol{\Sigma}^{\left(j\right)})^{-1}\notag\\&~~~~~~~~~~~~~~~~~~~~~~~~~~~~~~~~+\gamma^2(\gamma\boldsymbol{I}+\boldsymbol{\Sigma}^{\left(j\right)})^{-1}\boldsymbol{\tilde x}^{\left(j\right)}(\boldsymbol{\tilde x}^{\left(j\right)})^H(\gamma\boldsymbol{I}+\boldsymbol{\Sigma}^{\left(j\right)})^{-1}\big),\quad \forall j.
\end{align}Lemma \ref{lemma_appc} is thus proved.

\end{appendices}

\end{document}